\documentstyle[12pt]{article}
\makeatletter
\@addtoreset{equation}{section}
\renewcommand{\theequation}{\thesection.\arabic{equation}}
\makeatother
\makeatletter
\@addtoreset{equation}{subsection}

\def\R{\displaystyle \mathop{\bf R}}
\def\RR{\displaystyle \mathop{R}}

\def\G1{\displaystyle \mathop{G}}
\def\e1{\displaystyle \mathop{e}}
\def\U1{\displaystyle \mathop{U}}

\def\n1{\displaystyle \mathop{\nu}}
\def\ps1{\displaystyle \mathop{\Psi}}

\def\p1{\displaystyle \mathop{p}}
\def\J1{\displaystyle \mathop{J}}

\def\O1{\displaystyle \mathop{O}}

\def\hG{\displaystyle \mathop{\hat{G}}}
\def\gam{\displaystyle \mathop{\gamma}}

\def\pr{\displaystyle \mathop{\partial}}
\def\lpr{\displaystyle \mathop{\overleftarrow{\partial}}}
\def\lD1{\displaystyle \mathop{\overleftarrow{D}}}

\def\hgam{\displaystyle \mathop{\hat{\gamma}}}
\def\hpr{\displaystyle \mathop{\hat{\partial}}}
\def\Dlt{\displaystyle \mathop{\Delta}}
\def\hN{\displaystyle \mathop{\hat{N}}}
\def\he{\displaystyle \mathop{\hat{e}}}
\def\hp1{\displaystyle \mathop{\hat{p}}}
\def\hps{\displaystyle \mathop{\hat{\Psi}}}

\def\bp{\displaystyle \mathop{\bar{\Psi}}}

\def\S{\displaystyle \sum}
\def\IIn{\displaystyle \int}
\def\FFr{\displaystyle \frac}
\def\Lm{\displaystyle \lim}
\def\pd{\displaystyle \prod}
\def\hb{\displaystyle \mathop{\hat{b}}}
\def\F{\displaystyle \mathop{F}}
\def\sig1{\displaystyle \mathop{\sigma}}

\def\h1{\displaystyle \mathop{h}}
\def\H1{\displaystyle \mathop{H}}
\def\A1{\displaystyle \mathop{A}}

\def\D1{\displaystyle \mathop{D}}
\def\B1{\displaystyle \mathop{B}}
\def\L1{\displaystyle \mathop{L}}
\def\J1{\displaystyle \mathop{J}}
\def\A1{\displaystyle \mathop{A}}
\def\M1{\displaystyle \mathop{M}}
\def\g1{\displaystyle \mathop{g}}
\def\q1{\displaystyle \mathop{q}}
\def\x1{\displaystyle \mathop{x}}
\def\s1{\displaystyle \mathop{s}}
\begin{document}
\begin{center}
{\Large {\bf The Operator Manifold  Formalism. I}}
\end{center}
{\begin{flushright}
G.T.Ter-Kazarian\\
{\small Byurakan Astrophysical Observatory, Armenia 378433}\\
{\small E-mail:gago@bao.sci.am}\\
{\small December 20, 1998}
\end{flushright}
\begin{abstract}
The suggested operator manifold
formalism enables to develop an approach to the unification of the geometry
and the field theory.
The secondary quantization and differential geometric aspects are studied.
The former is equivalent to a configuration space wave mechanics incorporated
with geometric properties leading to the quantization of
geometry, different
in principle from earlier suggested schemes.
We show that the matrix elements of operator
tensors produce the Cartan's exterior forms,
define one parameter group of operator diffeomorphisms, 
consider the operator differential forms and their integration, 
also operator exterior differentiation.\\
We elaborate the formalism of operator
multimanifold yielding the multiworld geometry involving 
the spacetime continuum and internal worlds, where 
the subquarks are defined implying the Confinement and Gauge principles.
This formalism in Part II is used to develop further the microscopic
approach to the field theory.

\end{abstract} 

\section {Introduction}
\label {int}
A number of alternative approaches have been proposed
towards the unified gauge field theory, e.g. [1-7]. Each of them has
its own advantages and difficulties. The key problem is to find out
the mathematical structures enabling an insight to the concepts of
particle physics. An alternative approach is developed in our recent
work on the operator manifold formalism [8], elaborated in analogy of
secondary quantization incorporated with the geometric properties.
In the present paper we continue to study its background including the
rigorous definition of operator manifold and infer the matrix elements of 
field operators used for calculation of matrix elements of geometric 
objects.
We generalize this formalism via the  concept of operator multimanifold-
a multiworld geometry decomposed into the spacetime continuum and
internal worlds.\\
The operator manifold formalism has the following features:
1. It provides a natural unification of the geometry yielding  
Special and General Relativity principles and fermions serving as the 
basis for the subquarks (Part II).
2. They emerge in the geometry only in certain permissible combinations,
which utilizes the idea of Subcolour (Subquark) Confinement 
principle, and undergo the transformations yielding the internal
symmetries and Gauge principle (subsec.3.4,3.5).\\
This approach still should be considered as a preliminary one and  
numerous issues still remain to be solved. The only argument forcing us 
to consider it seriously is the fact that some important properties of 
particle physics can be derived naturally within this approach.

\renewcommand{\theequation}{\thesubsection.\arabic{equation}}
\section {Preliminaries}
\label{Prin}
This article is the continuation of [8], so we adopt its all ideas and 
notations, except the change in the order of vector and covector indices
to fit conventional notations used in differential 
geometry (subsec.2.4).  It is convenient to describe our approach 
in terms of manifold $G=\G1_{\eta}\oplus 
\G1_{u}$, (subsec.2.1) $ Dim \,G=12,\quad Dim\,\G1_{i}=6 \quad (i=\eta, u)$. 
But, one may readily return to conventional terms of 
Minkowski spacetime continuum (subsec.2.1). 
To be brief we often suppress the indices
without notice.\\
In Part II and further we deal with multiworld geometry, 
except for the
change of the concept of quark used in [8,9] as well as in Part I,
to subquark defined in the given
internal world.

\subsection{Operator Vector and Covector Fields}
\label{Vec}
Consider a curve $\lambda(t):R^{1}\rightarrow G$ passing through
a point $p=\lambda(0)\in G$ with tangent vector 
$\left.{\bf A}\right|_{\lambda(t)}$, where the $G$ is 12 dimensional
smooth differentiable manifold. The set $\{\zeta\}$ are local coordinates 
in open neighbourhood of $p\in \cal U$.
The 12 dimensional smooth vector field ${\bf A}_{p} ={\bf A}({\bf \zeta})$ 
belongs to the section of tangent bundle ${\bf T}_{p}$ at the point
$p({\bf \zeta})$. The one parameter
group of diffeomorphisms $A^{t}$ is given for the curve 
${\bf \zeta}(t)$ passing through point $p$ and ${\bf \zeta}(0)=
{\bf \zeta}_{p}, \quad \dot{{\bf \zeta}}(0)={\bf A}_{p}:\quad
d\,A^{t}_{p}({\bf A})=\left.\FFr{d}{d\,t}\right|_{t=0}A^{t}\left(
{\bf \zeta}(t)\right) = {\bf A}_{p}({\bf \zeta}).
$
Hence, 
$d\,A^{t}:{\bf T}(G)\rightarrow R$ $\left( 
{\bf T}(G) = \displaystyle \mathop{\bigcup}_{p({\bf \zeta})}
{\bf T}_{p}\right)$.
The 
$\{e_{(\lambda,\mu,\alpha)}=O_{\lambda,\mu}\otimes
\sigma_{\alpha}\} \subset G$
$(\lambda,\mu=1,2; \quad \alpha=1,2,3)$
is a set of linear independent $12$ unit vectors at the point $p$,
provided with the linear unit bipseudovectors $O_{\lambda,\mu}$
and the ordinary unit vectors $ \sigma_{\alpha}$ implying
$$<O_{\lambda,\mu},O_{\tau,\nu}>={}^{*}\delta_{\lambda,\tau}
{}^{*}\delta_{\mu,\nu}\quad
<\sigma_{\alpha}, \sigma_{\beta}>= \delta_{\alpha\beta},\quad
{}^{*}\delta=1-\delta,
$$
where $\delta$ is Kronecker symbol, the $\{ O_{\lambda,\mu}= 
O_{\lambda}\otimes  O_{\mu} \}$ is
the basis for tangent vectors of $2 \times 2$ dimensional linear 
pseudospace ${}^{*}{\bf R}^{4}={}^{*}{\bf R}^{2}\otimes
{}^{*}{\bf R}^{2}$, the $ \sigma_{\alpha}$ refers to three dimensional 
ordinary space ${\bf R}^{3}$. 
Henceforth we always let the first two subscripts in the parentheses
to denote the pseudovector components, while the third refers to the
ordinary vector components.
The metric on $G$ is 
$\hat{\bf g}:{\bf T}_{p}\otimes {\bf T}_{p}\rightarrow C^{\infty}(G)$ 
a section of conjugate vector bundle $S^{2}{\bf T}$.
Any vector ${\bf A}_{p}\in{\bf T}_{p}$ reads ${\bf A}=
e A$, provided with
components $A$ in the basis
$\{e\}$.
In holonomic coordinate basis $\left(\partial/\partial\,
\zeta\right)_{p}$ one gets 
$A=\left.\FFr{d\,
\zeta}{d\,t}\right|_{p}$ and
$\hat{g}=g d\zeta\otimes d\zeta $.
The manifold $G$ is decomposed as follows:
$$G={}^{*}{\bf R}^{2}
\otimes {}^{*}{\bf R}^{2} \otimes {\bf R}^{3}=\G1_{\eta}\oplus
\G1_{u}=\displaystyle\sum^{2}_{\lambda,\mu=1} 
\oplus {\bf R}^{3}_{\lambda \mu}=
{\R_{x}}^{3}\oplus 
{\R_{x_{0}}}^{3}\oplus
{\R_{u}}^{3}\oplus 
{\R_{u_{0}}}^{3}$$ with corresponding 
basis vectors  
${\e1_{i}}_{(\lambda\alpha)}={\O1_{i}}_{\lambda}\otimes 
\sigma_{\alpha}
\subset \G1_{i}$ $(\lambda =\pm,\quad 
i=\eta, u)$ of tangent sections, where 
$${\O1_{i}}_{+}=
\displaystyle \frac{1}{\sqrt{2}}(O_{1,1} +\varepsilon_{i} O_{2,1}),\quad
{\O1_{i}}_{-}=
\displaystyle \frac{1}{\sqrt{2}}(O_{1,2} +\varepsilon_{i} O_{2,2}),\quad
\varepsilon_{\eta}=1,\quad\varepsilon_{u}=-1.$$ Then
$<{\O1_{i}}_{\lambda},{\O1_{j}}_{\tau}>=
\varepsilon_{i}\delta_{ij}{}^{*}\delta_{\lambda \tau}$.
The $\G1_{\eta}$ is decomposed into three dimensional
ordinary and time  flat 
spaces $\G1_{\eta}=
{\R_{x}}^{3}\oplus {\R_{x_{0}}}^{3}$ with signatures 
$sgn({\R_{x}}^{3})=(+++)$ 
and $sgn({\R_{x_{0}}}^{3})=(---)$ (the same holds for $\G1_{u}$). 
The positive metric forms are defined
on manifolds $\G1_{i}:\quad $   
$\eta^{2}\in \G1_{\eta}, \quad
u^{2}\in \G1_{u}.$
The passage to Minkowski space is a further step as follows:
Since all directions in ${\R_{x_{0}}}^{3}$ are
equivalent, then by notion {\em time} one implies the projection of
time-coordinate on fixed arbitrary universal direction in ${\R_{x_{0}}}^{3}$.
By the reduction ${\R_{x_{0}}}^{3}\rightarrow {\R_{x_{0}}}^{1}$ the passage
$\G1_{\eta}\rightarrow M^{4}={\R_{x}}^{3}\oplus {\R_{x_{0}}}^{1}$ may be 
performed whenever it will be needed.
For more discussion of properties of
$G$ we refer to [10,11].\\
Unifying the geometry and particles into one framework
the operator manifold formalism is analogeous to the method of secondary 
quantization with appropriate expansion over the geometric objects.
For the secondary quantization of geometry,
first we substitute the basis elements by the 
creation and annihilation operators acting 
in the configuration space of occupation numbers.
Instead of pseudo vectors $O_{\lambda}$ we introduce the  
operators supplied by additional index ($r$) referring to the quantum 
numbers of corresponding state
\begin{equation}
\label {R24}
\begin{array}{ll}
\hat{O}^{r}_{1}=O^{r}_{1}\alpha_{1},\quad
\hat{O}^{r}_{2}=O^{r}_{2}{\alpha}_{2},\quad
\hat{O}_{r}^{\lambda}={}^{*}\delta^{\lambda\mu}\hat{O}^{r}_{\mu}=
{(\hat{O}^{r}_{\lambda})}^{+},\\ 
\{ \hat{O}^{r}_{\lambda},\hat{O}^{r'}_{\tau} \}=
\delta_{rr'}{}^{*}\delta_{\lambda\tau}I_{2}, 
\quad
<{O}^{r}_{\lambda},{O}^{r'}_{\tau}>= \delta_{rr'}{}^{*}\delta_{\lambda\tau},
\qquad I_{2}=\left( \matrix{
1 &0 \cr
0 &1\cr
}\right).
\end{array}
\end{equation}
The matrices ${\alpha}_{\lambda}$ satisfy the condition
$
\{ {\alpha}_{\lambda},{\alpha}_{\tau} \}={}^{*}\delta_
{\lambda\tau}I_{2},
$ where
$
{\alpha}^{\lambda}={}^{*}\delta^{\lambda\mu}
{\alpha}_{\mu}={({\alpha}_{\lambda})}^{+},
$
For example
${\alpha}_{1}=\left( \matrix{
0 &1 \cr
0 &0 \cr
}\right), \quad
{\alpha}_{2}=\left( \matrix{
0 &0 \cr
1 &0 \cr
}\right).$
Creation operator $\hat{O}^{r}_{1}$ generates one occupied state
$\mid 1>_{(0)}\equiv\mid 0,\ldots,1,\ldots>$ and the basis vector
$O^{r}_{1}$ with the quantum number $r$
through acting on nonoccupied vacuum state
$\mid 0>\equiv  \mid 0,0,\ldots>$:
$\quad
\hat{O}^{r}_{1}\mid 0>={O}^{r}_{1}\mid 1>.
$
Accordingly, the action of annihilation operator $\hat{O}^{r}_{2}$
on one occupied state yields the vacuum state and the basis vector
$O^{r}_{2}$:
$\quad
\hat{O}^{r}_{2}\mid 1>={O}^{r}_{2}\mid 0>.
$
So 
$\hat{O}^{r}_{1}\mid 1>=0, \\
\hat{O}^{r}_{2}\mid 0>=$0.
For instance, a matrix realization of the states
is
$\mid 0>\equiv\chi_{1}=\left( \matrix{
0 \cr
1\cr
}\right), \\
\mid 1>\equiv\chi_{2}=\left( \matrix{
1 \cr
0 \cr
}\right).$
The vacuum state reads
$\chi_{0}\equiv\mid 0>=\displaystyle\prod_{r=1}^{N}(\chi_{1})_{r}$.
The one occupied state is
$\chi_{r'}\equiv\mid 1>=(\chi_{2})_{r'}\displaystyle\prod_{r\neq r'}
(\chi_{1})_{r}$.
Instead of ordinary basis vectors we introduce
the operators
$\hat{\sigma}^{r}_{\alpha}\equiv\delta_{\alpha\beta\gamma}
\sigma^{r}_{\beta}\widetilde{\sigma}_{\gamma}$,
where $\widetilde{\sigma}_{\gamma}$ are Pauli's matrices, and
\begin{equation}
\label{R27}
<\sigma_{\alpha}^{r},\sigma_{\beta}^{r'}>=\delta_{rr'}\delta_{\alpha\beta},
\quad
\hat{\sigma}^{\alpha}_{r}=\delta^{\alpha\beta}\hat{\sigma}^{r}_{\beta}=
{(\hat{\sigma}_{\alpha}^{r})}^{+}=\hat{\sigma}_{\alpha}^{r},
\quad
\{\hat{\sigma}_{\alpha}^{r},\hat{\sigma}_{\beta}^{r'}\}=2
\delta_{rr'}\delta_{\alpha\beta}I_{2}.
\end{equation}
A matrix realization
of the vacuum state $\mid 0>\equiv{\varphi}_{1(\alpha)}$
and one occupied state $\mid 1_{(\alpha)}>
\equiv{\varphi}_{2(\alpha)}$ is as follows:
${\varphi}_{1(\alpha)}\equiv\chi_{1}, \quad
{\varphi}_{2(1)}=\left( \matrix{
1 \cr
0 \cr
}
\right), \quad
{\varphi}_{2(2)}=\left( \matrix{
-i \cr
0\cr
}
\right), \quad
{\varphi}_{2(3)}=\left( \matrix{
0 \cr
-1\cr
}
\right).$
Then
$$
{\hat{\sigma}}_{\alpha}^{r}\varphi_{1(\alpha)}=\sigma_
{\alpha}^{r}\varphi_{2(\alpha)}=(\sigma_{\alpha}^{r}\widetilde{\sigma}_
{\alpha})\varphi_{1(\alpha)},
\quad
{\hat{\sigma}}_{\alpha}^{r}\varphi_{2(\alpha)}=\sigma_
{\alpha}^{r}\varphi_{1(\alpha)}=(\sigma_{\alpha}^{r}\widetilde{\sigma}_
{\alpha})\varphi_{2(\alpha)}.
$$
Hence, the single eigenvalue
$(\sigma_{\alpha}^{r}\widetilde{\sigma}_{\alpha})$
associates with different $\varphi_{\lambda(\alpha)}$,
namely it is degenerated with 
degeneracy degree equal 2. So, among quantum numbers $r$
there is also the quantum number of the half integer spin 
$\vec{\sigma}$ 
$(\sigma_{3}=\FFr{1}{2}s,\quad s=\pm1)$.
As it will be seen, this consequently {\em gives rise to the spins of
particles}.
One occupied state reads
$\varphi_{r'(\alpha)}={(\varphi_{2(\alpha)})}_{r'}\displaystyle
\prod_{r\neq r'}{(\chi_{1})}_{r}$.
Next we introduce the operator
$$
{\hat{\gamma}}^{r}_{(\lambda,\mu,\alpha)}\equiv{\hat{O}}^{r_{1}}_{\lambda}
\otimes{\hat{O}}^{r_{2}}_{\mu}\otimes{\hat{\sigma}}^{r_{3}}_{\alpha}$$
and the state vector
$$
\chi_{\lambda,\mu,\tau(\alpha)}\equiv\mid\lambda,\mu,\tau(\alpha)>=
\chi_{\lambda}\otimes\chi_{\mu}\otimes\varphi_{\tau(\alpha)}, 
$$
where $\lambda,\mu,\tau,\nu=
1,2;\quad \alpha,\beta=1,2,3$ and $r\equiv (r_{1},r_{2},r_{3})$.
Omitting two valuedness of state vector we apply
$\mid\lambda,\tau,\delta(\beta)>\equiv\mid\lambda,\tau>$,
and remember that always the summation 
must be extended over the double degeneracy of the spin states $(s=\pm 1)$.
One infers the explicit form of corresponding
matrix elements
\begin{equation}
\label{R211}
<\lambda,\mu\mid{\hat{\gamma}}^{r}_{(\tau,\nu,\alpha)}\mid \tau,\nu>=
{}^{*}\delta_{\lambda\tau}{}^{*}\delta_{\mu\nu}
e^{r}_{(\tau,\nu,\alpha)},\quad
<\tau,\nu\mid{\hat{\gamma}}_{r}^{(\tau,\nu,\alpha)}\mid\lambda,\mu >=
{}^{*}\delta_{\lambda\tau}{}^{*}\delta_{\mu\nu}
e_{r}^{(\tau,\nu,\alpha)}.
\end{equation}
for given $\lambda,\mu.$
The operators of occupation numbers are
\begin{equation}
\label{R212}
{\hN_{1}}^{rr'}_{\alpha\beta}=
{\hat{\gamma}}^{r}_{(1,1,\alpha)}{\hat{\gamma}}^{r'}_{(2,2,\beta)},
\quad
{\hN_{2}}^{rr'}_{\alpha\beta}=
{\hat{\gamma}}^{r}_{(2,1,\alpha)}{\hat{\gamma}}^{r'}_{(1,2,\beta)},
\end{equation}
with the expectation values implying Pauli's exclusion principle
\begin{equation}
\label{R214}
\begin{array}{ll}
<2,2\mid{\hN_{1}}^{rr'}_{\alpha\beta}\mid 2,2>=\delta_{rr'}
\delta_{\alpha\beta},
\quad
<1,2\mid{\hN_{2}}^{rr'}_{\alpha\beta}\mid 1,2>=\delta_{rr'}\delta_
{\alpha\beta},\\ 
<1,1\mid{\hN_{1}}^{rr'}_{\alpha\beta}\mid 1,1>=0,\qquad
<2,1\mid{\hN_{2}}^{rr'}_{\alpha\beta}\mid 2,1>=0.
\end{array}
\end{equation}
The set of operators $\{{\hat{\gamma}}^{r}\}$
is the basis for tangent operator vectors 
$\hat{\Phi}(\zeta)={\hat{\gamma}}^{r}
\Phi_{r}(\zeta)$ of the 12 dimensional flat operator manifold $\hat{G}$,
where we introduce the vector function
belonging to the ordinary class of functions of $C^{\infty}$ smoothness 
defined on the manifold $G$:
$\quad \Phi_{r}^{(\lambda,\mu,\alpha)}(\zeta)=
\zeta^{(\lambda,\mu,\alpha)} \Phi_{r}^{\lambda,\mu}(\zeta),\quad
\zeta \in G$.
But a set of operators $\{{\hat{\gamma}}_{r}\}$
is a dual basis for
operator covectors
$\bar{\hat{\Phi}}(\zeta)={\hat{\gamma}}_{r}
\Phi^{r}(\zeta)$, where
$\Phi^{r}=
{\bar{\Phi}}_{r}$ (charge conjugated).
One gets
\begin{equation}
\label{R215}
<\lambda,\mu\mid\hat{\Phi}(\zeta)\bar{\hat{\Phi}}(\zeta)\mid \lambda,\mu>=
{}^{*}\delta_{\lambda\tau}{}^{*}\delta_{\mu\nu}
\Phi_{r}^{(\tau,\nu,\alpha)}(\zeta)\Phi^{r}_{(\tau,\nu,\alpha)}(\zeta),
\end{equation}
for given $\lambda,\mu.$
Considering the  state vectors
\begin{equation}
\label{R216}
\begin{array}{lll}
\chi^{0}(\nu_{1},\nu_{2},\nu_{3},\nu_{4})=
\mid 1,1>^{\nu_{1}}\cdot\mid 1,2>^{\nu_{2}}\cdot
\mid 2,1>^{\nu_{3}}\cdot\mid 2,2>^{\nu_{4}},\\
\nu_{i}= \left\{ \begin{array}{ll}
                   1   & \mbox{if $\nu=\nu_{i}$}\quad  \mbox{for some $i$,} \\
                   0   & \mbox{otherwise},
                   \end{array}
\right. 
\\
\mid\chi_{-}(1)>= \chi^{0}(1,0,0,0),\quad   
\mid\chi_{+}(1)>=\chi^{0}(0,0,0,1),\quad
<\chi_{\pm}(\lambda)\mid\chi_{\pm}(\mu)>=\delta_{\lambda\mu}, \\
\mid\chi_{-}(2)>  = \chi^{0}(0,0,1,0),\quad   
\quad
\mid\chi_{+}(2)>= \chi^{0}(0,1,0,0),\quad
<\chi_{\pm}(\lambda)\mid\chi_{\mp}(\mu)>=0,
\end{array}
\end{equation}
provided
$
<\chi_{\pm}\mid A\mid \chi_{\pm}>\equiv
\S_{\lambda}<\chi_{\pm}(\lambda)\mid A\mid \chi_{\pm}(\lambda)>, 
$
we get the matrix elements
\begin{equation}
\label{R217}
\begin{array}{l}
<\chi_{+}\mid\hat{\Phi}(\zeta)\bar{\hat{\Phi}}(\zeta)\mid\chi_{+}>
\equiv\Phi^{2}_{+}(\zeta)=\quad
\Phi_{r}^{(\lambda,1,\alpha)}(\zeta)\Phi^{r}_{(\lambda,1,\alpha)}(\zeta),\\ 
<\chi_{-}\mid\hat{\Phi}(\zeta)\bar{\hat{\Phi}}(\zeta)\mid\chi_{-}>
\equiv\Phi^{2}_{-}(\zeta)=
\Phi_{r}^{(\lambda,2,\alpha)}(\zeta)\Phi^{r}_{(\lambda,2,\alpha)}(\zeta).
\end{array}
\end{equation}
The basis $\{{\hat{\gamma}}^{r}\}$
is decomposed into
$\{ {\hgam_{i}}^{r} \}\quad
(\lambda=\pm;\quad\alpha=1,2,3;\quad i=\eta,u):\quad$
${\hgam_{i}}^{r}_{(+\alpha)}=\FFr{1}{\sqrt{2}}
(\gamma^{r}_{(1,1\alpha)}+\varepsilon_{i}
\gamma^{r}_{(2,1\alpha)}),
\quad
{\hgam_{i}}^{r}_{(-\alpha)}=\FFr{1}{\sqrt{2}}
(\gamma^{r}_{(1,2\alpha)}+\varepsilon_{i}
\gamma^{r}_{(2,2\alpha)}).$ 
The expansions of operator vectors $\hps_{i}\in\hG_{i}$ and 
operator covectors  $\bar{\hps_{i}}$ are written 
$\hps_{i}={\hgam_{i}}^{r}{\ps1_{i}}_{r},
\quad\bar{\hps_{i}}=
{\hgam_{i}}_{r}{\ps1_{i}}^{r},$
where the vector functions of $C^{\infty}$ smoothness are 
defined on the manifolds $\G1_{i}$: 
\begin{equation}
\label {R23}
{\ps1_{\eta} }_{r}^{(\pm\alpha)}(\eta,p_{\eta})=\eta^{(\pm\alpha)}
{\ps1_{\eta} }_{r}^{\pm}(\eta,p_{\eta}),\quad 
{\ps1_{u}}_{r}^{(\pm\alpha)}(u,p_{u})=u^{(\pm\alpha)}
{\ps1_{u}}_{r}^{\pm}(u,p_{u}).
\end{equation}
It is assumed that the probability of finding the vector function 
in the state $r$ with given sixvector of coordinate ($\eta$ or $u$) and 
momentum ($p_{\eta}$ or $p_{u}$) is determined by the square of its state 
wave function ${\ps1_{\eta} }_{r}^{\pm}(\eta,p_{\eta}),$ or 
${\ps1_{u}}_{r}^{\pm}(u,p_{u})$. 
Due to the spin states, the ${\ps1_{i}}_{r}^{\pm}$ is the Fermi field 
of the positive and negative frequencies 
${\ps1_{i}}^{\pm}_{r}={\ps1_{i}}^{r}_{\pm p}$.

\subsection {Realization of the Flat Manifold $G$} 
\label {quant}
The bispinor $\Psi(\zeta)$ defined on manifold $G=\G1_{\eta}
\oplus\G1_{u}$ is written  $\Psi(\zeta)=\ps1_{\eta}(\eta)\ps1_{u}(u)$,
where the $\ps1_{i}$ is a bispinor defined on 
the manifold $\G1_{i}.$
The free state of $i$-type fermion with definite values of momentum
$p_{i}$ and spin projection $s$ is described by means of plane waves,
($\hbar=1, c=1$)[8]:
$
{\ps1_{\eta}}_{p_{\eta}}(\eta)={\left({\FFr{m}{E_{\eta}}}
\right)} ^{1/2}
\U1_{\eta}(p_{\eta},s)e^{-ip_{\eta}\eta} 
$,
etc  where $E_{i}\equiv {\p1_{i}}_{0}=\mid \vec{\p1_{i}}_{0}\mid ,
\quad
{\p1_{i}}_{0\alpha}=\FFr{1}{\sqrt{2}}({\p1_{i}}_{(+\alpha)}+
{\p1_{i}}_{(-\alpha)}),\quad
\vec{\p1_{i}}=\FFr{1}{\sqrt{2}}(\vec{\p1_{i}}_{+}-
\vec{\p1_{i}}_{-}),
\quad
p^{2}_{\eta}=E^{2}_{\eta}-{\vec{p}}^{2}_{\eta}=p^{2}_{u}=
E^{2}_{u}-{\vec{p}}^{2}_{u}=m^{2}$.
We consider also the solutions of negative
frequencies.
For the spinors the useful relations of orthogonality and
completeness hold.
We make use of localized wave packets constructed by
means of superposition of plane wave solutions furnished by creation
and annihilation operators in agreement with Pauli's principle 
$$
{\hps_{i}}=\S_{\pm s}\IIn\frac{d^{3}p_{i}}{{(2\pi)}^{3/2}}
\left( {\hgam_{i}}_{(+\alpha)}{\ps1_{i}}^{(+\alpha)}+
{\hgam_{i}}_{(-\alpha)}{\ps1_{i}}^{(-\alpha)}\right), 
$$
etc, where the summation is extended over all dummy indices.
The matrix element of the anticommutator of expansion coefficients reads
\begin{equation}   
\label{R35}
<\chi_{-}\mid \{ {\hgam_{i}}^{(+\alpha)}(p_{i},s),
{\hgam_{j}}_{(+\beta)}(p'_{j},s')\}\mid\chi_{-}>=
\varepsilon_{i}\delta_{ij}\delta_{ss'}\delta_{\alpha\beta}\delta^{(3)}
({\vec{p}}_{i}- {\vec{p'}}_{i}).
\end{equation}
We also consider wave packets of operator vector
fields $\hat{\Phi}(\zeta)$.
In this manner we get the important relation
\begin{equation}
\label{R37}
\begin{array}{l}
\S_{\lambda=\pm}<\chi_{\lambda}\mid\hat{\Phi}(\zeta)
\bar{\hat{\Phi}}(\zeta)\mid
\chi_{\lambda}>= 
\S_{\lambda=\pm}<\chi_{\lambda}\mid
\bar{\hat{\Phi}}(\zeta)\hat{\Phi}(\zeta)\mid\chi_{\lambda}>= \\ 
=-i\Lm_{\zeta\rightarrow\zeta'}(\zeta\zeta')\G1_{\zeta}(\zeta-\zeta')
=-i\left[ \Lm_{\eta\rightarrow\eta'}(\eta\eta')\G1_{\eta}(\eta-\eta')-
\Lm_{u\rightarrow u'}(uu')\G1_{u}(u-u')\right],
\end{array}
\end{equation}
where the Green's function 
$
\G1_{i}(i-i')=-(i\hpr_{i}+m)\Dlt_{i}(i-i')
$
is used,
provided with the invariant singular functions 
$\Dlt_{i}(i-i')$ $(i=\eta,u)$.
Realization of the manifold $G$ is due to the  
constraint imposed upon the matrix element eq.(2.2.2)
which is, as a geometric object, required to be finite
\begin{equation}
\label{R919}
\S_{\lambda=\pm}<\chi_{\lambda}\mid\hat{\Phi}(\zeta)
\bar{\hat{\Phi}}(\zeta)\mid
\chi_{\lambda}>=
\zeta^{2}{\G1_{\zeta}}_{F}(0) < \infty.
\end{equation}
Thereto [8] (see eq.(3.1.3))
\begin{equation}
\label{R920}
\begin{array}{l}
{\G1_{\zeta}}_{F}(0)={\G1_{\eta}}_{F}(0)={\G1_{u}}_{F}(0)=\\
= \Lm_{u\rightarrow u'}\left[ -i\S_{{\vec{p}}_{u}}{\ps1_{u}}_{{p}_{u}}(u)
{\bp_{u}}_{{p}_{u}}(u')\theta (u_{0}-u'_{0})+
i\S_{{\vec{p}}_{u}}{\bp_{u}}_{{-p}_{u}}(u'){\ps1_{u}}_{{-p}_{u}}(u)
\theta (u'_{0}-u_{0}) \right].
\end{array}
\end{equation}
The ${\G1_{\zeta}}_{F},{\G1_{\eta}}_{F}$ and ${\G1_{u}}_{F}$ are
causal Green's functions 
characterized by the boundary condition that only positive frequency
occur for $\eta_{0}>0\quad(u_{0}>0)$, only negative for
$\eta_{0}<0\quad(u_{0}<0)$. Here $\eta_{0}=\mid \vec{\eta}_{0}\mid $, 
$\eta_{0\alpha}=\FFr{1}{\sqrt{2}}
(\eta_{(+\alpha)}+\eta_{(-\alpha)})$ and the same holds for $u_{0}$.
Then, satisfying the condition eq.(2.2.3) a length of each vector
${\bf \zeta}=e\zeta\in G$
(see eq.(2.2.2)) compulsory should be equaled zero
\begin{equation}
\label{R921}
\zeta^{2}=\eta^{2}-u^{2}=0.
\end{equation}
Thus the requirement eq.(2.2.3)
provided by eq.(2.2.4) yields the realization of the flat
manifold $G$}, which subsequently leads to Minkowski flat
space $M^{4}$ (subsec.2.1), where according to eq.(2.2.5)
the Relativity principle holds
$$
\left. d\,\eta^{2}\right|_{6\rightarrow 4}\equiv d\,s^{2}=
d\,u^{2}=inv.
$$

\subsection{Mathematical Background: Field Aspect}
\label {appe}
\renewcommand {\theequation}{\thesubsection.\arabic {equation}}
The quantum field theory of operator manifold 
is equivalent to configuration space wave mechanics employing the 
antisymmetric state functions incorporated with geometric 
properties of corresponding objects [12]. 
In this subsection we reach
to rigorous definition of concept of operator manifold
$\hat{G}$, construct the explicit forms of wave state 
functions and calculate the matrix elements of field operators.\\
Suppose that the $i$-th fermion is found in the state $r_{i}$ 
with the vector function $\Phi_{r_{i}}^{(\lambda_{i},\mu_{i},\alpha_{i})}=
\zeta_{r_{i}}^{(\lambda_{i},\mu_{i},\alpha_{i})}
\Phi_{r_{i}}^{\lambda_{i},\mu_{i}}(\zeta_{r_{i}})$ (eq.(2.1.6)) and 
\begin{center}
$\zeta_{r_{i}^{\lambda_{i},\mu_{i}}}=
\S^{3}_{\alpha_{i}=1}
e^{r_{i}}_{(\lambda_{i},\mu_{i},\alpha_{i})}
\zeta_{r_{i}}^{(\lambda_{i},\mu_{i},\alpha_{i})}$, \quad
$\zeta_{r_{i}}=\S^{2}_{\lambda_{i},\mu_{i}=1}
\zeta_{r_{i}^{\lambda_{i},\mu_{i}}}\in \widetilde{\cal U}_{r_{i}}$, 
\end{center}
the $\widetilde{\cal U}
_{r_{i}}$
is the open neighbourhood of the point $\zeta_{r_{i}};$ the $r_{i}$ 
implies a set $\left(r_{i}^{11},r_{i}^{12},r_{i}^{21},r_{i}^{22}\right)$.
Let the ${\cal H}^{(1)}$ is a Hilbert space used for quantum mechanical 
description of one particle, namely ${\cal H}^{(1)}$ is a finite or infinite
dimensional complex space provided with scalar product $(\Phi,\Psi),$
which is linear with respect to $\Psi$ and antilinear to $\Phi$.  
The ${\cal H}^{(1)}$ is complete in norm 
$|\Phi|=(\Phi,\Phi)^{1/2}$,
i.e. each fundamental sequence $\{\Phi_{n}\}$ of vectors of ${\cal H}^{(1)}$
is converged by norm on ${\cal H}^{(1)}$. 
One particle state function is written 
$\Phi^{(1)}_{r_{i}}=\displaystyle \prod^{2}_{\lambda_{i},\mu_{i}=1}
\Phi^{(1)}_{r_{i}^{\lambda_{i},\mu_{i}}}\in {\cal H}^{(1)}_{r_{i}}$, where
${\cal H}^{(1)}_{r_{i}}=
\displaystyle \prod^{2}_{\lambda_{i},\mu_{i}=1}\otimes
{\cal H}^{(1)}_{r_{i}^{\lambda_{i},\mu_{i}}}$. Define
\begin{equation}
\label{R39}
\widetilde{\Phi}^{(1)}=\zeta_{i}\Phi^{(1)}_{r_{i}}\in 
\widetilde{G}^{(1)}_{r_{i}}=\widetilde{\cal U}^{(1)}_{r_{i}}\otimes
{\cal H}^{(1)}_{r_{i}}.
\end{equation}
For description of n particle system we introduce Hilbert space
\begin{equation}
\label{R319}
{\bar{\cal H}}^{(n)}_{(r_{1},\ldots,r_{n})}=
{\cal H}^{(1)}_{r_{1}}\otimes\cdots\otimes
{\cal H}^{(1)}_{r_{n}}
\end{equation}
by considering all sequences
\begin{equation}
\label{R310}
\Phi^{(n)}_{(r_{1},\ldots,r_{n})}=
\{\Phi^{(1)}_{r_{1}},\ldots,\Phi^{(1)}_{r_{n}}\}=
\Phi^{(1)}_{r_{1}}\otimes \cdots \otimes \Phi^{(1)}_{r_{n}},
\end{equation}
where $\Phi^{(1)}_{r_{i}}\in {\cal H}^{(1)}_{r_{i}}$ provided, 
as usual, with the scalar product
\begin{equation}
\label{R312}
(\Phi^{(n)}_{(r_{1},\ldots,r_{n})},\Psi^{(n)}_{(r_{1},\ldots,r_{n})})=
\prod^{n}_{i=1}(\Phi^{(1)}_{r_{i}},\Psi^{(1)}_{r_{i}}).
\end{equation}
We consider the space 
${\cal H}^{(n)}_{(r_{1},\ldots,r_{n})}$ of all limited linear combinations 
of eq.(2.3.2) and continue by linearity the scalar product eq.(2.3.4)
on ${\cal H}^{(n)}_{(r_{1},\ldots,r_{n})}$. The wave function
$\Phi^{(n)}_{(r_{1},\ldots,r_{n})} 
\in {\cal H}^{(n)}_{(r_{1},\ldots,r_{n})}$ 
must be antisymmetrized over its arguments. We distinguish the
antisymmetric part ${}^{A}{\bar{\cal H}}^{(n)}$ of Hilbert space
${\bar{\cal H}}^{(n)}$ by considering the functions
\begin{equation}
\label{R313}
{}^{A}\Phi^{(n)}_{(r_{1},\ldots,r_{n})}=\FFr{1}{\sqrt{n!}}\S_{\sigma\in S(n)}
sgn(\sigma)\Phi^{(n)}_{\sigma(r_{1},\ldots,r_{n})}.
\end{equation}
The summation is extended over all permutations of indices
$(r_{1}^{\lambda\mu},\ldots, r_{n}^{\lambda\mu})$ of the integers
$1,2,\ldots,n,$ where the antisymmetrical eigenfunctions are sums
of the same terms with alternating signs in dependence of a parity
$sgn(\sigma)$ of transposition. One continues the reflection 
$\Phi^{(n)}\rightarrow {}^{A}\Phi^{(n)}$ by linearity on ${\cal H}^{(n)}$, 
which is limited and 
enables the expansion by linearity on ${}^{A}{\bar{\cal H}}^{(n)}$.
The region of values of this
reflection is a ${}^{A}{\bar{\cal H}}^{(n)}$, namely an antisymmetrized
tensor product of $n$ identical samples of ${}{\cal H}^{(1)}$.
We introduce
\begin{equation}
\label{R314}
\begin{array}{l}
{}^{A}{\widetilde{\Phi}}^{(n)}_{(r_{1},\ldots,r_{n})}=
\FFr{1}{\sqrt{n!}}\S_{\sigma\in S(n)}
sgn(\sigma){\widetilde {\Phi}}^{(n)}_{\sigma(r_{1},\ldots,r_{n})}=\\ 
=\FFr{1}{\sqrt{n!}}\S_{\sigma\in S(n)}
sgn(\sigma){\widetilde{\Phi}}^{(1)}_{r_{1}}\otimes \cdots \otimes 
{\widetilde{\Phi}}^{(1)}_{r_{n}} \in 
{}^{A}{\widetilde{G}}^{(n)}_{(r_{1},\ldots,r_{n})}
=\widetilde{\cal U}^{(n)}_{(r_{1},\ldots,r_{n})}\otimes
{}^{A}{\hat{\cal H}}^{(n)}_{(r_{1},\ldots,r_{n})}.
\end{array}
\end{equation}
and consider a set ${}^{A}\widetilde{\cal F}$ of all sequences
$
{}^{A}{\widetilde{\Phi}}=\{ {}^{A}{\widetilde{\Phi}}^{(0)},
{}^{A}{\widetilde{\Phi}}^{(1)}\ldots ,
{}^{A}{\widetilde{\Phi}}^{(n)}\ldots \},
$
with a finite number of nonzero elements. Therewith, the set 
${}^{A}\cal F$
$
{}^{A}\Phi=\{ {}^{A}\Phi^{(0)},
{}^{A}\Phi^{(1)}\ldots ,
{}^{A}\Phi^{(n)}\ldots \}
$
is provided with the structure of Hilbert subspace employing
the composition rules
\begin{equation}
\label{R317}
\begin{array}{l}
{}^{A}(\lambda\Phi +\mu\Psi)^{(n)}=
\lambda {}^{A}\Phi^{(n)} + \mu {}^{A}\Psi^{(n)}, \quad 
\forall \lambda, \mu \in C,\\ 
({}^{A}\Phi,{}^{A}\Psi)=\S^{\infty}_{n=0}({}^{A}\Phi^{(n)},{}^{A}\Psi^{(n)}).
\end{array}
\end{equation}
The wave manifold ${\cal G}$ stems from the
${}^{A}\widetilde{\cal F}$ is due to the expansion by metric 
induced as a scalar product on ${}^{A}\cal F$
\begin{equation}
\label{R318}
{\cal G}=\S^{\infty}_{n=0}{\cal G}^{(n)}=
\S^{\infty}_{n=0}\left(\widetilde{\cal U}^{(n)}\otimes {}^{A}
\bar{\cal H}^{(n)}\right).
\end{equation}
The creation ${\hat{\gamma}}_{r}$ and
annihilation ${\hat{\gamma}}^{r}$ operators for each
${}{\cal H}^{(1)}$ can be defined as follows:
one must modify the basis operators in order to provide
an anticommutation in both the same and
different states
\begin{equation}
\label{R319}
{\hat{\gamma}}_{(\lambda,\mu,\alpha)}^{r}\rightarrow
{\hat{\gamma}}_{(\lambda,\mu,\alpha)}^{r}\eta_{r}^{\lambda\mu},
\quad
{(\eta_{r}^{\lambda\mu})}^{+}=\eta_{r}^{\lambda\mu},
\end{equation}
for given $\lambda,\mu,\alpha$,
where $\eta_{r}$ is a diagonal operator in the space 
of occupation numbers,
while, at $r_{i}<r_{j}$ one gets
$
{\hat{\gamma}}^{r_{i}}\eta_{r_{j}}=
-\eta_{r_{j}}{\hat{\gamma}}^{r_{i}},
\quad
{\hat{\gamma}}^{r_{j}}\eta_{r_{i}}=
\eta_{r_{i}}{\hat{\gamma}}^{r_{j}}.
$
The operators of corresponding occupation numbers (for given 
$\lambda,\mu,\alpha)$ are
$
{\hat{N}}_{r}={\hat{\gamma}}^{r}
{\hat{\gamma}}_{r}$.
Since the diagonal operators $(1-2{\hat{N}}_{r})$ 
anticommute with the
${\hat{\gamma}}^{r}$, then
$
\eta_{r_{i}}=\prod_{r =1}^{r_{i}-1}(1-2{\hat{N}}_{r}),
$
where
\begin{equation}
\label{R325}
\eta_{r_{i}}^{11}\Phi(n_{1},\ldots,n_{N};0;0;0)=\pd_{r =1}^{r_{i}-1}
(-1)^{n_{r}}\Phi(n_{1},\ldots,n_{N};0;0;0),
\end{equation}
etc.
Here the occupation numbers $n_{r}(m_{r},q_{r},t_{r})$ are introduced, 
which refer to the $r$-th states corresponding to operators
${\hat{\gamma}}_{(1,1,\alpha)}^{r}$, etc
either empty ($n_{r},\ldots,t_{r}=0$) or occupied
($n_{r},\ldots,t_{r}=1$).
To save writing we abbreviate the modified operators by the
same symbols. For example,
acting on free state $\mid 0>_{r_{i}}$ 
the creation operator ${\hat{\gamma}}_{r_{i}}$
yields the one occupied state
$\mid 1>_{r_{i}}$ with the phase $+$ or $-$ depending of parity of the number
of quanta in the states $ r < r_{i}$.
Modified operators satisfy the same anticommutation relations
of the basis operators (subsec.2.1). 
It is convenient to make use of notation
${\hat{\gamma}}^{(\lambda,\mu,\alpha)}_{r}\equiv
{e}^{(\lambda,\mu,\alpha)}_{r}{\hat{b}}^{\lambda\mu}_{(r\alpha)},$,
and abbreviate the pair of indices $(r\alpha)$ by the single symbol $r$. 
Then for each $\Phi \in {}^{A}{\cal H}^{(n)}$
and any vector $f \in {\cal H}^{(1)}$ the operators
$\hat{b}(f)$ and
$\hat{b}^{*}(f)$ imply
\begin{equation}
\label{R327}
\begin{array}{l}
\hat{b}(f)\Phi=\FFr{1}{\sqrt{(n-1)!}}\S_{\sigma\in S(n)}
sgn(\sigma)\left( f\,\Phi^{(1)}_{\sigma{(1)}} \right)
\Phi^{(1)}_{\sigma{(2)}}\otimes\cdots\otimes \Phi^{(1)}_{\sigma{(n)}},\\
\hat{b}^{*}(f)\Phi=\FFr{1}{\sqrt{(n+1)!}}\S_{\sigma\in S(n+1)}
sgn(\sigma)\Phi^{(1)}_{\sigma{(0)}}\otimes
\Phi^{(1)}_{\sigma{(1)}}\otimes\cdots\otimes \Phi^{(1)}_{\sigma{(n)}},
\end{array}
\end{equation}
where $\Phi^{(1)}_{(0)}\equiv f$. One continues the 
$\hat{b}(f)$ and $\hat{b}^{*}(f)$ by linearity to linear reflections, 
which are denoted by the same symbols acting  respectively from 
${}^{A}{\cal H}^{(n)}$ into ${}^{A}{\cal H}^{(n-1)}$ or 
${}^{A}{\cal H}^{(n+1)}$.
They are limited over the values $\sqrt{n}|f|$ and $\sqrt{(n+1)}|f|$
and can be expanded by continuation up to the reflections acting 
from ${}^{A}{\bar{\cal H}}^{(n)}$
into ${}^{A}{\bar{\cal H}}^{(n-1)}$ or ${}^{A}{\bar{\cal H}}^{(n+1)}$.
Finally, they must be continued by linearity up to the linear
operators acting from ${}^{A}{\cal F}$ into ${}^{A}{\cal F}$
defined on the same closed region in
${}^{A}{\bar{\cal H}}^{(n)}$, namely in ${}^{A}{\cal F}$, which is 
invariant with respect to reflections
$\hat{b}(f)$ and $\hat{b}^{*}(f)$. Hence, at $f_{i}, g_{i}\in 
{\cal H}^{(1)}$ $(i=1,\ldots,n; j=1,\ldots,m)$ all polynomials over 
$\{\hat{b}^{*}(f_{i}) \}$ and $\{\hat{b}(g_{j}) \}$ are completely defined 
on ${}^{A}{\cal F}$. While, for given $\lambda,\mu$, one has
\begin{equation}
\label{R328}
\begin{array}{l}
<\lambda,\mu\mid\{ {\hat{b}}^{\lambda\mu}_{r}(f),{\hat{b}}_{\lambda\mu}
^{r'}(g)\}\mid \lambda,\mu>=
\delta^{r'}_{r}.
\end{array}
\end{equation}
The mean values $<\varphi; {\hat{b}}_{r}^{\lambda\mu}(f)
{\hat{b}}^{r}_{\lambda\mu}(f)>$ calculated at fixed $\lambda,\mu$ for
any element $\Phi \in {}^{A}{\cal F}$ equal to mean values of the symmetric 
operator of occupation number in terms of
${\hat{N}}^{r}={\hat{b}}_{r}(f)
{\hat{b}}^{r}(f)$, with a wave function $f$ in the 
state described by $\Phi$. 
Here, as usual, it is denoted $<\varphi; A\Phi> = Tr \,
P_{\varphi}A = (\Phi,A\Phi)$ for each vector $\Phi \in {\cal H}$ with
$|\Phi |=1$, while the $P_{\varphi}$ is projecting operator onto one
dimensional space $\{\lambda \Phi\left.\right| \lambda\in C\}$ generated by
$\Phi$. Therewith, the probability of transition $\varphi \rightarrow \psi$  
is given $Pr\{\varphi\left.\right|\psi\} =\left| (\psi,\varphi)\right|^{2}$.
The linear operator $A$ is defined on the elements 
of linear manifold ${\cal D}(A)$ of ${\cal H}$ taking the values in
${\cal H}$. The ${\cal D}(A)$ is an everywhere closed region of definition
of $A$, namely the closure of ${\cal D}(A)$ by the norm given in
${\cal H}$ coincides with ${\cal H}$. Meanwhile, the ${\cal D}(A)$ is 
included in ${\cal D}(A^{*})$ and  $A$ coincides with the reduction of 
$A^{*}$ on ${\cal D}(A)$, because the ${\cal D}(A)$ is a symmetric operator
and the linear operator $A^{*}$ is maximal conjugated to $A$. That is,
any operator $A'$ conjugated to $A$ - $(\Psi, A'\Phi)=(A'\Psi, \Phi)$ 
at all $\Phi \in {\cal D}(A)$ and $\Psi \in {\cal D}(A')$ coincides 
with the reduction of $A^{*}$ on some linear manifold ${\cal D}(A')$
included in ${\cal D}(A^{*})$. Thus, the operator $A^{**}$ is closed 
symmetric expansion of operator $A$, namely it is a closure of $A$.
Self conjugated operator $A$ (the closure of which is self conjugated too)
allows only one self conjugated expansion $A^{**}$. Thus, 
self conjugated closure $\hat{N}$ of operator 
$\S_{i=1}^{\infty}
\hat{b}^{*}(f_{i})\hat{b}(f_{i})$, where $\{f_{i}\left.\right| i=1,\ldots, 
n\}$ is an arbitrary orthogonal basis on ${\cal H}^{(1)}$, is regarded
as the operator of occupation number. For the vector $\chi^{0} \in 
{}^{A}{\cal F}$ and $\chi^{0(n)}=\delta_{0n}$ one gets $<\chi^{0(n)},
\hat{N}(f)>=0$ for all $f \in {\cal H}^{(1)}$.
So, the $\chi^{0}$ is a vector of vacuum state:
$\hat{b}(f)\chi^{0}=0$ for all $f \in {\cal H}^{(1)}$. If
$f =\{ f_{i}\left.\right| i=1,2,\ldots\}$ is an arbitrary orthogonal basis 
on ${\cal H}^{(1)}$, then due to irreducibility of operators 
$\hat{b}^{*}(f_{i})\left.\right| f_{i} \in f$, the ${}^{A}{\cal H}$
includes the $0$ and whole space ${}^{A}{\cal H}$ as invariant subspaces 
with respect to all $\hat{b}^{*}(f)$. 
To define the 12 dimensional operator manifold 
$\hat{G}$ we consider a set $\hat{{\cal F}}$ of all sequences
$\hat{\Phi}=\{\hat{\Phi}^{(0)},\hat{\Phi}^{(1)},\ldots,\hat{\Phi}^{(n)},
\ldots \}$ with a finite number of nonzero elements provided
\begin{equation}
\label{R329}
\begin{array}{l}
\hat{\Phi}^{(n)}_{(r_{1},\ldots,r_{n})}=\hat{\Phi}^{(1)}_{r_{1}}
\otimes \cdots \otimes \hat{\Phi}^{(1)}_{r_{n}}\in \hat{G}^{(n)},\quad
\hat{\Phi}^{(1)}_{r_{i}}=\hat{\zeta}_{r_{i}}\Phi^{(1)}_{r_{i}}\in
\hat{G}^{(1)}_{i}=\hat{\cal U}^{(1)}_{i}\otimes {\cal H}^{(1)}_{i},\\
\hat{\zeta}_{r_{i}}\equiv \S^{3}_{\alpha_{i}=1}
{\hat{\gamma}}^{r_{i}}_{(\lambda_{i},\mu_{i},\alpha_{i})}
\zeta_{r_{i}}^{(\lambda_{i},\mu_{i},\alpha_{i})}\in
\hat{\cal U}^{(1)}_{r_{i}}, \quad \hat{G}^{(n)}=\hat{\cal U}^{(n)}
\otimes \bar{\cal H}^{(n)}, \quad
\hat{\cal U}^{(n)}_{(r_{1},\ldots,r_{n})}=\hat{\cal U}^{(1)}_{r_{1}},
\otimes\cdots\otimes \hat{\cal U}^{(1)}_{r_{n}}.
\end{array}
\end{equation}
Then on the analogy of eq.(2.3.8) the operator manifold $\hat{G}$
ensues
\begin{equation}
\label{R330}
\hat{G}=\S^{\infty}_{n=0}\hat{G}^{(n)}=
\S^{\infty}_{n=0}\left(\hat{\cal U}^{(n)}\otimes {\bar{\cal H}}^{(n)}\right).
\end{equation}
To define the 
secondary quantized form of one particle observable $A$ on $\cal H$, 
following [13]
let consider a set of identical samples $\hat{\cal H}_{i}$ of
one particle space ${\cal H}^{(1)}$ and operators $A_{i}$ acting in them.
To each closed linear operator $A^{(1)}$ in ${\cal H}^{(1)}$ with 
everywhere closed region of definition ${\cal D}(A^{(1)})$  
following operators are corresponded:
\begin{equation}
\label{R331}
\begin{array}{l}
A^{(n)}_{1}=A^{(1)}\otimes I \otimes\cdots \otimes I,\\
\ldots\ldots\ldots\ldots\ldots\ldots\ldots\ldots \\
A^{(n)}_{n}=I\otimes I \otimes\cdots \otimes A^{(1)}.
\end{array}
\end{equation}
Their sum $\S^{n}_{j=1}A^{(n)}_{j}$ is given on the intersection of regions of
definition of operator terms including a linear manifold
${\cal D}(A^{(1)})\otimes\cdots \otimes {\cal D}(A^{(n)})$ being closed
in $\hat{\cal H}^{(n)}$. While, the $A^{(n)}$ is a minimal closed expansion 
of this sum with ${\cal D}(A^{(n)})$. One considers a linear manifold
${\cal D}(\Omega(A))$ in ${\cal H}=\S^{\infty}_{n=0}\hat{\cal H}^{(n)}$
defined as a set of all vectors $\Psi \in {\cal H}$ such as 
$\Psi^{(n)} \in {\cal D}(A^{(n)})$ and $\S^{\infty}_{n=0}\left|
A^{(n)}\Psi^{(n)}\right|^{2} < \infty$ . The manifold
${\cal D}(\Omega(A))$ is closed in ${\cal H}$. On this manifold one defines 
a closed linear operator $\Omega(A)$ acting as $\Omega(A)^{(n)}=
A^{(n)}\Psi^{(n)}$, namely $\Omega(A)\Phi = \S^{\infty}_{n=0}
A^{(n)}\Psi^{(n)}$, while the $\Omega(A)$ is self conjugated  operator
with everywhere closed region of definition. We suppose that the vector
$\Phi^{(n)} \in {\cal H}^{(n)}$ is in the form eq.(2.3.3),
where $\Phi_{i}\in {\cal D}(A)$. Then $<\varphi^{(n)};A^{(n)}>=
\S^{n}_{i=1}<\varphi_{i};A>$, which enables the expansion by 
continuing onto ${\cal D}(A)$. Thus $A^{(n)}$ is the $n$ particle 
observable corresponding to one particle observable $A$. So
$<\varphi;\Omega(A)>=
\S^{\infty}_{n=0}<\varphi^{(n)};A^{(n)}>$ for any $\Phi_{i}\in 
{\cal D}\left(\Omega(A)\right)$. While, the $\Omega(A)$ reflects
${}^{A}{\cal D}={\cal D}\left(\Omega(A)\right)\frown {}^{A}{\cal H}$
into ${}^{A}{\cal H}$.
The reduction of $\Omega(A)$ on ${}^{A}{\cal H}$ is self conjugated
in the region ${}^{A}{\cal D}$, because of the fact that ${}^{A}{\cal H}$
is a closed subspace of ${\cal H}$. Hence, the $\Omega(A)$ is a secondary
quantized form of one particle observable $A$ on ${\cal H}$.\\
The vacuum state reads eq.(2.1.7) with the normalization requirement
\begin{equation}
\label{R332}
<\chi^{0}(\nu'_{1},\nu'_{2},\nu'_{3},\nu'_{4})\mid
\chi^{0}(\nu_{1},\nu_{2},\nu_{3},\nu_{4})>=\prod_{i=1}^{4}
\delta_{\nu_{i}\nu'_{i}}.
\end{equation}
The state vectors
\begin{equation}
\label{R333}
\begin{array}{l}
\chi({\{n_{r}\}}^{N}_{1};{\{m_{r}\}}^{M}_{1};
{\{q_{r}\}}^{Q}_{1};{\{t_{r}\}}^{T}_{1};
{\{\nu_{r}\}}^{4}_{1})=
{(\hat{b}_{N}^{11})}^{n_{\scriptscriptstyle N}}\cdots
{(\hat{b}_{1}^{11})}^{n_{1}}\cdot\\ 
\cdot {(\hat{b}_{M}^{12})}^{m_{\scriptscriptstyle M}}\cdots
{(\hat{b}_{1}^{12})}^{m_{1}}\cdot
{(\hat{b}_{Q}^{21})}^{q_{\scriptscriptstyle Q}}\cdots
{(\hat{b}_{1}^{21})}^{q_{1}}\cdot
\cdot{(\hat{b}_{T}^{22})}^{t_{\scriptscriptstyle T}}\cdots
{(\hat{b}_{1}^{22})}^{t_{1}}
\chi^{0}(\nu_{1},\nu_{2},\nu_{3},\nu_{4}),
\end{array}
\end{equation}
where ${\{n_{r}\}}^{N}_{1}=n_{1},\ldots,n_{N}$, etc are the 
eigenfunctions
of modified operators. They form a whole set of orthogonal vectors
\begin{equation}
\label{R334}
\begin{array}{l}
<\chi,\chi'>
=\displaystyle \prod_{r=1}^{N}\delta_{n_{r}n'_{r}}\cdot
\displaystyle \prod_{r=1}^{M}\delta_{m_{r}m'_{r}}\cdot
\displaystyle \prod_{r=1}^{Q}\delta_{q_{r}q'_{r}}\cdot
\displaystyle \prod_{r=1}^{T}\delta_{t_{r}t'_{r}}\cdot
\displaystyle \prod_{r=1}^{4}\delta_{\nu_{r}\nu'_{r}}.
\end{array}
\end{equation}
Considering an arbitrary superposition
\begin{equation}
\label{R335}
\chi=
\S^{1}_{
a={\{n_{r}\}}^{N}_{1},{\{m_{r}\}}^{M}_{1},
{\{q_{r}\}}^{Q}_{1},{\{t_{r}\}}^{T}_{1}=0
}
c'(a)\, \chi(a),
\end{equation}
the coefficients $c'$ of expansion are the corresponding amplitudes
of probabilities.
Taking into account eq.(2.3.12), the nonvanishing matrix elements of
operators $\hat{b}^{11}_{r_{k}}$ and $\hat{b}_{11}^{r_{k}}$ read
\begin{equation}
\label{R337}
\begin{array}{l}
<\chi({\{n'_{r}\}}^{N}_{1};0;0;0;1,0,0,0)\left|   \right.
\hat{b}^{11}_{r_{k}}
\chi({\{n_{r}\}}^{N}_{1};0;0;0;1,0,0,0)>=\\
\\
=<1,1\mid \hat{b}_{11}^{r'_{1}}\cdots\hat{b}_{11}^{r'_{n}}\cdot
\hat{b}^{11}_{r_{k}}\cdot
\hat{b}^{11}_{r_{n}}\cdots\hat{b}^{11}_{r_{1}}\mid 1,1>=\\
\\
= \left\{ \begin{array}{ll}
(-1)^{n'-k'}  & \mbox{if $n_{r}=n'_{r}$ for $r\neq r_{k}$ and $n_{r_{k}}=0;n'_{r_{k}}=1$}, \\
0           & \mbox{otherwise},
\end{array}  \right.   \\

<\chi({\{n'_{r}\}}^{N}_{1};0;0;0;1,0,0,0)\left|   \right.
\hat{b}_{11}^{r_{k}}
\chi({\{n_{r}\}}^{N}_{1};0;0;0;1,0,0,0)>=\\
\\
=<1,1\mid \hat{b}_{11}^{r'_{1}}\cdots\hat{b}_{11}^{r'_{n}}\cdot
\hat{b}_{11}^{r_{k}}\cdot
\hat{b}^{11}_{r_{n}}\cdots\hat{b}^{11}_{r_{1}}\mid 1,1>=\\
\\
= \left\{ \begin{array}{ll}
(-1)^{n-k}  & \mbox{if $n_{r}=n'_{r}$ for $r\neq r_{k}$ and $n'_{r_{k}}=0;n_{r_{k}}=1$}, \\
0           & \mbox{otherwise},
\end{array} \right.
\end{array}
\end{equation}
where one denotes
$
n=\S_{r=1}^{N}n_{r}, \quad n'=\S_{r=1}^{N}n'_{r},
$
the $r_{k}$ and $r'_{k}$ are $k$-th and $k'$-th terms of regulated sets of
$\{r_{1},\ldots ,r_{n}\} \quad (r_{1}<r_{2}<\cdots <r_{n})$ and
$\{r'_{1},\ldots ,r'_{n}\} \quad (r'_{1}<r'_{2}<\cdots <r'_{n})$,
respectively.
Continuing along this line we get a whole set of explicit forms of
matrix elements of the rest of operators
$\hat{b}_{r_{k}}$ and $\hat{b}^{r_{k}}$.
Thus 
\begin{equation}
\label{R339}
\S_{\{\nu_{r}\}=0}^{1}<\chi^{0}\mid\hat{\Phi}(\zeta)\mid\chi>= 
\S_{r=1}^{N}c'_{n_{r}}e_{(1,1,\alpha)}^{n_{r}}
{\Phi}^{(1,1,\alpha)}_{n_{r}}
+\cdots,
\end{equation}
provided
\begin{equation}
\label{R340}
c'_{n_{r}}\equiv\delta_{1n_{r}}c'(0,\ldots, n_{r},\ldots,0;0;0;0),
\cdots
\end{equation}
Hereinafter we change the notation 
\begin{equation}
\label{R341}
\begin{array}{l}
\bar{c}(r^{11})=c'_{n_{r}},\quad \bar{c}(r^{21})=c'_{q_{r}},
\quad N_{11}=N, \quad N_{21}=Q, \\ 
\bar{c}(r^{12})=c'_{m_{r}}, \quad \bar{c}(r^{22})=c'_{t_{r}},
\quad N_{12}=M, \quad N_{22}=T,
\end{array}
\end{equation}
and make use of
\begin{equation}
\label{R342}
F_{r^{\lambda\mu}}= \S_{\alpha}
e_{(\lambda,\mu,\alpha)}^{r^{\lambda\mu}}
{\Phi}^{(\lambda,\mu,\alpha)}_{r^{\lambda\mu}},
\quad
\S_{\{\nu_{r}\}=0}^{1}<\chi^{0}\mid\hat{A}\mid\chi>
\equiv<\chi^{0}\parallel\hat{A}\parallel\chi>,
\end{equation}
The matrix elements of operator vector and covector fields
take the forms
\begin{equation}
\label{R343}
\begin{array}{l}
<\chi^{0}\parallel\hat{\Phi}(\zeta)\parallel\chi>=
\S_{\lambda\mu=1}^{2}\S_{r^{\lambda\mu}=1}^{N_{\lambda\mu}}
\bar{c}(r^{\lambda\mu})F_{r^{\lambda\mu}}(\zeta),\\
<\chi\parallel\bar{\hat{\Phi}}(\zeta)\parallel\chi^{0}>=
\S_{\lambda\mu=1}^{2}\S_{r^{\lambda\mu}=1}^{N_{\lambda\mu}}
{\bar{c}}^{*}(r^{\lambda\mu})F^{r^{\lambda\mu}}(\zeta).
\end{array}
\end{equation}
In the following we shall use:
$\left\{\S_{\lambda\mu}^{2} \right\}_{1}^{n}
\equiv \S_{\lambda_{1}\mu_{1}}^{2}\ldots\S_{\lambda_{n}\mu_{n}}^{2},\quad$
$r^{\lambda\mu}_{i}\equiv r^{\lambda_{i}\mu_{i}}$ and
$ 
\bar{c}(r^{11}_{1},\ldots,r^{11}_{n})=c'(n_{1},\ldots,n_{N};0;0;0),
$ etc.
The anticommutation relations ensue
\begin{equation}
\label{R346}
<\chi_{-}\mid \{ {\hb_{i}}^{+}_{r},{\hb_{i}}_{+}^{r'} \}\mid\chi_{-}>=
<\chi_{+}\mid \{ {\hb_{i}}^{-}_{r},{\hb_{i}}_{-}^{r'} \}\mid\chi_{+}>=
{\delta}^{r'}_{r},
\end{equation}
provided 
${\hgam_{i}}^{(\lambda\alpha)}_{r}=
{\he_{i}}^{(\lambda\alpha)}_{r}{\hb_{i}}^{\lambda}_{(r\alpha)}, \quad
(r\alpha)\rightarrow r.$
The state functions
\begin{equation}
\label{R347}
\chi=
{({\hb_{\eta}}_{N}^{+})}^{n_{\scriptscriptstyle N}}\cdots
{({\hb_{\eta}}_{1}^{+})}^{n_{1}}
\cdot
{({\hb_{\eta}}_{M}^{-})}^{m_{\scriptscriptstyle M}}\cdots
{({\hb_{\eta}}_{1}^{-})}^{m_{1}}\cdot
{({\hb_{u}}_{Q}^{+})}^{q_{\scriptscriptstyle Q}}\cdots
{({\hb_{u}}_{1}^{+})}^{q_{1}}
\cdot{({\hb_{u}}_{T}^{-})}^{t_{\scriptscriptstyle T}}\cdots
{({\hb_{u}}_{1}^{-})}^{t_{1}}\cdot
\chi_{-}(\lambda)\chi_{+}(\mu),
\end{equation}
form a whole set of orthogonal
eigenfunctions of corresponding operators of occupation numbers
${\hN_{i}}^{\lambda}_{r}={\hb_{i}}_{r}^{\lambda}{\hb_{i}}^{r}_{\lambda}$
with the expectation values 0,1.

\subsection {Differential Geometric Aspect}
\label {diff}
The set of operators $\{ \hat{\gamma}^{r} \}$ is
the basis for all operator vectors of tangent section ${\hat{\bf T}}
_{\Phi_{p}}$
of principle bundle with the base $\hat{G}$ at the point 
${\bf \Phi}_{p}=\left. {\bf \Phi}({\bf \zeta}(t))\right|_{t=0} \in
\hat{G}$.
The smooth field of tangent operator vector 
$\hat{\bf A}({\bf \Phi}({\bf \zeta}))$ is a class of equivalency of the 
curves ${\bf f}({\bf \Phi}({\bf \zeta}))$,
${\bf f}({\bf \Phi}({\bf \zeta}(0)))={\bf \Phi}_{p}$.
While, the operator differential $\hat{d}\,A^{t}_{p}$ of the flux
$A^{t}_{p}:\hat{G} \rightarrow \hat{G}$ at the point
${\bf \Phi}_{p}$ with the velocity fields 
$\hat{\bf A}({\bf \Phi}({\bf \zeta}))$ is defined by one parameter
group of operator diffeomorphisms given for the curve 
${\bf \Phi}({\bf \zeta}(t)):R^{1} \rightarrow \hat{G}$, provided
${\bf \Phi}({\bf \zeta}(0))={\bf \Phi}_{p}$ and
$\widehat{\dot{{\bf \Phi}}}({\bf \zeta}(0))=\hat{\bf A}_{p}$
\begin{equation}
\label{R41}
\hat{d}\,A^{t}_{p}({\bf A})=\left.\FFr{\hat{d}}{d\,t}\right|_{t=0}
A^{t}({\bf \Phi}({\bf \zeta}(t)))=\hat{\bf A}({\bf \Phi}({\bf \zeta}))=
\hat{\gamma}^{r}A_{p},
\end{equation}
where the $\{ A_{p}\}$ are the components of $\hat{\bf A}$ in the basis 
$\{ \hat{\gamma}^{r}\}$. According to eq.(2.4.1),
in holonomic coordinate basis 
$ \hat{\gamma}^{r} \rightarrow \left(
\hat{\partial}\left. \right/ \partial \Phi_{r}
({\bf \zeta}(t))\right)_{p}$ one gets
$
A_{p}=\left. \FFr{\partial \Phi_{r}}{\partial \zeta_{r}}
\FFr{d\,\zeta_{r}}{d\,t}\right|_{p}.
$
\\
The operator tensor $\hat{\bf T}$ of 
$\widehat{(n,0)}$-type at the point ${\bf \Phi}_{p}$ is a linear
function of the space
$
\hat{\bf T}^{n}_{0}=\underbrace{\hat{\bf T}_{\Phi_{p}}\otimes\cdots \otimes 
\hat{\bf T}_{\Phi_{p}}}_{n},
$
where $\otimes $ stands for tensor product.
It enables a correspondence between the element $(\hat{\bf A}_{1},\ldots,
\hat{\bf A}_{n})$ of $\hat{\bf T}^{n}_{0}$ and the number $T(\hat{\bf A}_{1},
\ldots,\hat{\bf A}_{n})$, provided by linearity. 
An explicit form of matrix element of 
operator tensor reads
\begin{equation}
\label{R46}
\begin{array}{l}
\FFr{1}{\sqrt{n!}}<\chi^{0}\parallel\hat{\Phi}({\zeta}_{1})\otimes
\cdots \otimes\hat{\Phi}({\zeta}_{n})\parallel\chi>=\\
=\left\{\S_{\lambda\mu=1}^{2}\right\}_{1}^{n}
\S_{r_{1}^{\lambda\mu},\ldots ,r_{n}^{\lambda\mu}=1}^{N_{\lambda\mu}}
\bar{c}(r_{1}^{\lambda\mu},\ldots ,r_{n}^{\lambda\mu})
F_{r_{1}^{\lambda\mu}}(\zeta_{1})\wedge\cdots\wedge
F_{r_{n}^{\lambda\mu}}(\zeta_{n}),
\end{array}
\end{equation}
where $\wedge$ stands for exterior product. 
So, constructing matrix 
elements of operator tensors of $\hat{G}$ one produces the 
Cartan's exterior forms on wave manifold $\cal{G}$. 
Whence, the matrix elements of symmetric 
operator tensors equal zero. \\
The linear operator form of $1$ degree $\hat{\bf \omega}^{1}$ is a linear
operator valued function on $\hat{\bf T}_{\Phi_{p}}$, namely 
$\hat{\bf \omega}^{1}(\hat{\bf A}_{p}):\hat{\bf T}_{\Phi_{p}} 
\rightarrow \hat{R}$,
where $\hat{\bf A}_{p} \in \hat{\bf T}_{\Phi_{p}}$, and the operator
$\hat{\bf \omega}^{1}(\hat{\bf A})=<\hat{\bf \omega}^{1},{\bf A}> 
\in \hat{R}$ is corresponded to $\hat{\bf A}_{p}$ at the point 
${\bf \Phi}_{p}$, provided, according to eq.(2.3.26), with
\begin{equation}
\label{R47}
<\chi \| \hat{\bf \omega }^{1} \|
\chi^{0} >=
\S_{\lambda,\mu=1}^{2}
\S_{r^{\lambda\mu}=1}^{N_{\lambda\mu}} \hat{c}^{*}(r^{\lambda\mu})
{\bf \omega }^{1}_{r^{\lambda\mu}},
\end{equation}
where ${\bf \omega}^{1}_{r^{\lambda\mu}}=
e_{r^{\lambda\mu}}^{(\lambda,\mu,\alpha)}
\omega^{r^{\lambda\mu}}_{(\lambda,\mu,\alpha)}$, the 
$<{\bf \omega}^{1}_{r^{\lambda\mu}},{\bf A}>=
\omega ^{1}_{r^{\lambda\mu}}({\bf A})$ is a linear form on 
${\bf T}_{p}$, and 
\begin{equation}
\label{R48}
\begin{array}{l}
\hat{\bf \omega}^{1}(\lambda_{1}\hat{\bf A}_{1} + 
\lambda_{2}\hat{\bf A}_{2})=\lambda_{1}\hat{\bf \omega}^{1}(\hat{\bf A}_{1})+
\lambda_{2}\hat{\bf \omega}^{1}(\hat{\bf A}_{2}),\\ 
\forall \lambda_{1},\lambda_{2} \in R,\quad
\hat{\bf A}_{1}, \hat{\bf A}_{2} \in \hat{\bf T}_{\Phi_{p}}.
\end{array}
\end{equation}
The set of all linear operator forms defined at the point ${\bf \Phi}_{p}$
fill the operator vector space $\hat{\bf T}_{\Phi_{p}}^{*}$ 
dual to $\hat{\bf T}_{\Phi_{p}}$. While, the
$\{ \hat{\gamma}_{r} \}$ serves as a basis for them.
The operator $n$ form is defined as the exterior product of operator 
1 forms
\begin{equation}
\label{R49}
\hat{\bf \omega}^{n}(\hat{\bf A}_{1},\ldots,\hat{\bf A}_{n})=
\left(\hat{\bf \omega}^{1}_{1}\wedge \cdots \wedge 
\hat{\bf \omega}^{1}_{n}\right)
\left(\hat{\bf A}_{1},\ldots,\hat{\bf A}_{n}\right) =\\ 
\left\|
\begin{array}{lll}
\hat{\bf \omega}^{1}_{1}(\hat{\bf A}_{1}) 
\cdots\cdots 
&\hat{\bf \omega}^{1}_{n}(\hat{\bf A}_{1}\\
\vdots  &\vdots\\
\hat{\bf \omega}^{1}_{1}(\hat{\bf A}_{n}) 
\cdots\cdots 
&\hat{\bf \omega}^{1}_{n}(\hat{\bf A}_{n})
\end{array}
\right \| .
\end{equation}
Here as well as for the rest of this section 
we abbreviate the set of indices $(\lambda_{i},\mu_{i},\alpha_{i})$
by the single symbol $i$. 
If $\{\hat{\gamma}_{i}^{r_{i}}\}$ and $\{\hat{\gamma}^{i}_{r_{i}}\}$
are dual basises respectively in $\hat{\bf T}_{\Phi_{p}}$ and
$\hat{\bf T}_{\Phi_{p}}^{*}$, then the
$\{\hat{\gamma}_{1}^{r_{1}}\otimes\cdots\otimes\hat{\gamma}_{p}^{r_{p}}
\otimes\hat{\gamma}^{1}_{s_{1}}\otimes\cdots\otimes\hat{\gamma}^{q}_{s_{q}}\}$
will be the basis in operator space
$
\hat{\bf T}^{p}_{q}=\underbrace{\hat{\bf T}_{\Phi_{p}}\otimes\cdots \otimes 
\hat{\bf T}_{\Phi_{p}}}_{p}
\otimes\underbrace{\hat{\bf T}^{*}_{\Phi_{p}}\otimes\cdots \otimes 
\hat{\bf T}^{*}_{\Phi_{p}}}_{q}.
$
Any operator tensor $\hat{\bf T} \in \hat{\bf T}^{p}_{q}({\bf \Phi}_{p})$
can be written
$$
\hat{\bf T}=T^{i_{1}\cdots i_{p}}_{j_{1}\cdots j_{q}}
\left(r_{1},\ldots,r_{p},s_{1},\ldots,s_{q}\right)
\hat{\gamma}_{i_{1}}^{r_{1}}\otimes\cdots\otimes\hat{\gamma}_{i_{p}}^{r_{p}}
\otimes\hat{\gamma}^{j_{1}}_{s_{1}}\otimes\cdots\otimes
\hat{\gamma}^{j_{q}}_{s_{q}},
$$
where 
$
T^{i_{1}\cdots i_{p}}_{j_{1}\cdots j_{q}} 
\left(r_{1},\ldots,r_{p},s_{1},\ldots,s_{q}\right)=
T\left( 
\hat{\gamma}^{i_{1}}_{r_{1}}\otimes\cdots\otimes\hat{\gamma}^{i_{p}}_{r_{p}}
\otimes\hat{\gamma}_{j_{1}}^{s_{1}}\otimes\cdots\otimes
\hat{\gamma}_{j_{q}}^{s_{q}}
\right)
$
are the components of $\hat{\bf T}$ in 
$\{\hat{\gamma}_{i}^{r_{i}}\}$ and $\{\hat{\gamma}^{i}_{r_{i}}\}$.
For any function $f:{\bf R}^{n}\rightarrow {\bf R}^{n}$ of the
ordinary class of functions of $C^{\infty}$ smoothness on $\hat{G}$,
according to eq.(2.4.1), one defines an operator differential 
$$
<\hat{d}\,f,\hat{\bf A}>=(Af)^{\wedge},
$$
by means of
smooth reflection 
$$\hat{d}\,f:\hat{\bf T}\left(\hat{G}\right)
\rightarrow \hat{R}\quad \left(\hat{\bf T}\left(\hat{G}\right)=
\displaystyle \bigcup_{\Phi_{p}}\hat{\bf T}_{\Phi_{p}}\right),
$$ 
where 
$
<\chi\| B^{\wedge}\|\chi^{0}>=\S_{\lambda,\mu=1}^{2}
\S_{r^{\lambda\mu}=1}^{N_{\lambda\mu}} \hat{c}^{*}(r^{\lambda\mu})
B(r^{\lambda\mu})$
(see eq.(2.4.1)). Then
\begin{equation}
\label{R416}
<\chi\|\hat{d}\,f,\hat{\bf A} \|\chi^{0}>=\S_{\lambda,\mu=1}^{2}
\S_{r^{\lambda\mu}=1}^{N_{\lambda\mu}} \hat{c}^{*}(r^{\lambda\mu})
<d\,f,{\bf A}>_{r^{\lambda\mu}}
=\S_{\lambda,\mu=1}^{2}
\S_{r^{\lambda\mu}=1}^{N_{\lambda\mu}} \hat{c}^{*}(r^{\lambda\mu})
({\bf A}\,f)_{r^{\lambda\mu}}.
\end{equation}
In coordinate basis
$
<d\, \Phi^{\widehat{\imath}}, \hat{\partial}\left.\right/ \partial 
\Phi^{j}>=\FFr{\partial \Phi^{\widehat{\imath}}}{\partial \Phi^{j}}=
\delta^{{\widehat{\imath}}}_{j},
$
provided $d\, \Phi^{\widehat{\imath}}\equiv \hat{d}\Phi^{i}$ and
$
<\chi\|\delta^{{\widehat{\imath}}}_{j} \|\chi^{0}>=\S_{\lambda,\mu=1}^{2}
\S_{r^{\lambda\mu}=1}^{N_{\lambda\mu}} \hat{c}^{*}(r^{\lambda\mu})
\delta^{i}_{j}.
$
We define the differential operator
$n$ form $\left.{\bf \hat{\omega}}^{n}\right|_{\Phi_{p}}$ at the point
${\bf \Phi}_{p} \in \hat{G}$ as the exterior operator $n$ form
on tangent operator space $\hat{\bf T}_{\Phi_{p}}$ of tangent
operator vectors $\hat{\bf A}_{1},\ldots,\hat{\bf A}_{n}$.
That is, if the $\wedge \hat{\bf T}^{*}_{\Phi_{p}}
\left(\hat{G}\right)$ means the exterior algebra on
$\hat{\bf T}^{*}_{\Phi_{p}}\left(\hat{G}\right)$, then
operator $n$ form $\left.{\bf \hat{\omega}}^{n}\right|_{\Phi_{p}}$
is an element of $n$-th degree out of $\wedge \hat{\bf T}^{*}_{\Phi_{p}}$
depending upon the point ${\bf \Phi}_{p} \in \hat{G}$.
Hence ${\bf \hat{\omega}}^{n}=\displaystyle \bigcup_{\Phi_{p}}
\left.{\bf \hat{\omega}}^{n}\right|_{\Phi_{p}}$. Any differential operator
$n$ form of dual operator space
$\underbrace{\hat{\bf T}^{*}_{\Phi_{p}}\otimes\cdots \otimes 
\hat{\bf T}^{*}_{\Phi_{p}}}_{n}$ may be written 
\begin{equation}
\label{R419}
{\bf \hat{\omega}}^{n}=\S_{i_{1}<\cdots<i_{n}}\alpha
_{i_{1} \cdots i_{n}} (\Phi)d\,\Phi^{\widehat{\imath}_{1}}\wedge\cdots\wedge
d\,\Phi^{\widehat{\imath}_{n}},
\end{equation}
provided by the smooth differentiable functions $\alpha
_{i_{1} \cdots i_{n}}(\Phi)\in C^{\infty}$ and basis
$
d\,\Phi^{\widehat{\imath}_{1}}\wedge\cdots\wedge
d\,\Phi^{\widehat{\imath}_{n}}=
\S_{\sigma\in S_{n}}sgn(\sigma)
\gamma^{\sigma(\widehat{\imath}_{1}}\otimes\cdots\otimes
\gamma^{\widehat{\imath}_{n})}.
$
Any antisymmetric operator tensor of $\widehat{(0,n)}$ type
reads 
\begin{equation}
\label{R421}
\hat{\bf T}^{*}=T_{i_{1} \cdots i_{n}}\gamma^{\widehat{\imath}_{1}}
\otimes\cdots\otimes \gamma^{\widehat{\imath}_{n}}
=\S_{i_{1}<\cdots<i_{n}}
T_{i_{1} \cdots i_{n}} d\,\Phi^{\widehat{\imath_{1}}}\wedge\cdots\wedge
d\,\Phi^{\widehat{\imath_{n}}}.
\end{equation}
Let the $\hat{\cal D}_{1}$ and $\hat{\cal D}_{2}$ are two compact 
convex parallelepipeds in oriented $n$ dimensional operator space
$\hat{\bf R}^{n}$ and the $f:\hat{\cal D}_{1}\rightarrow \hat{\cal D}_{2}$
is differentiable reflection of interior of $\hat{\cal D}_{1}$ into
$\hat{\cal D}_{2}$ retaining an orientation, namely for any function
$\varphi \in C^{\infty}$ defined on $\hat{\cal D}_{2}$ it holds
$\varphi\circ f\in C^{\infty}$ and $f^{*}\varphi\left({\bf \Phi}_{p}\right)=
\varphi\left( f \left({\bf \Phi}_{p} \right)\right)$, where $f^{*}$
is an image of function $\varphi\left( f \left({\bf \Phi}_{p} \right)\right)$
on $\hat{\cal D}_{1}$ at the point ${\bf \Phi}_{p}$. Hence, the function
$f$ induces a linear reflection $\hat{d}\, f:\hat{\bf T}\left(
\hat{\cal D}_{1}\right)\rightarrow \hat{\bf T}\left(
\hat{\cal D}_{2}\right)$ as an operator differential of $f$ implying
$\hat{d}\,f\left(\hat{\bf A}_{p} \right)\varphi=\hat{\bf A}_{p}
(\varphi\circ f)$ for any operator vector $\hat{\bf A}_{p} \in
\hat{\bf T}_{\Phi_{p}}$ and for any function $\varphi \in C^{\infty}$ defined
in the neighbourhood of ${\bf \Phi'}_{p}=f\left({\bf \Phi}_{p}\right)$.
If the function $f$ is given in the form ${\Phi'}^{i}={\Phi'}^{i}\left(
\Phi_{p}\right)$ and
$\hat{\bf A}_{p}=\left(A^{i}\left.\hat{\partial}\right/ \partial\Phi^{i}
\right)_{p}$, then in terms of local coordinates one gets
$
\left(\hat{d}\,f\right)\hat{\bf A}_{p}=A^{i}\left(\FFr{\partial{\Phi'}^{j}}
{\partial\Phi^{i}}\right)_{p}
\left(\FFr{\hat{\partial}}
{\partial{\Phi'}^{j}}\right)_{p'}.
$
So, if $f_{1}:\hat{\cal D}_{1}\rightarrow \hat{\cal D}_{2}$ and
$f_{2}:\hat{\cal D}_{2}\rightarrow \hat{\cal D}_{3}$ then
$\hat{d}\,\left(f_{2}\circ f_{1}\right)=\hat{d}\,f_{2}\circ \hat{d}\,
f_{1}$. The differentiable reflection 
$f:\hat{\cal D}_{1}\rightarrow \hat{\cal D}_{2}$ induces the reflection
$\hat{f}^{*}:\hat{\bf T}^{*}\left(\hat{\cal D}_{2}\right)\rightarrow 
\hat{\bf T}^{*}\left(\hat{\cal D}_{1}\right)$ conjugated to $\hat{f}_{*}$.
The latter is the operator differential of $f$, while
\begin{equation}
\label{R423}
<\hat{f}^{*}\hat{\omega'}^{1},\hat{\bf A}>_{\Phi_{p}}=
\left.<\hat{\omega'}^{1},\hat{f}_{*}\hat{\bf A}>\right|_{f\left(
\Phi_{p}\right)},
\end{equation}
where $\left.\hat{\bf A}\right|_{f\left(\Phi_{p}\right)}=
\left(\hat{d}\,f\right)\hat{\bf A}_{p}$ and $\hat{\omega'}^{1}\in 
\left.\hat{\bf T}^{*}\right|_{f\left(\Phi_{p}\right)}$. Hence
$
\hat{f}^{*}\left(\hat{d}\,\varphi\right)=\hat{d}\,
\left(\hat{f}^{*}\varphi\right)
$
and
\begin{equation}
\label{R426}
\begin{array}{l}
\hat{f}^{*}\left.T\left(\hat{\bf A}_{1},\ldots,
\hat{\bf A}_{n}\right)\right|_{\Phi_{p}}=
\left.T\left(\hat{f}_{*}\hat{\bf A}_{1},\ldots,\hat{f}_{*}
\hat{\bf A}_{n}\right)\right|_{f\left(\Phi_{p}\right)},\\
\left.T\left(\hat{f}^{*}\hat{\bf \omega}^{1}_{1},\ldots,
\hat{f}^{*}\hat{\bf \omega}^{1}_{n}\right)\right|_{\Phi_{p}}=
\hat{f}_{*}\left.T\left(\hat{\bf \omega}^{1}_{1},\ldots,\hat{f}^{*}
\hat{\bf \omega}^{1}_{n}\right)\right|
_{f\left(\Phi_{p}\right)}.
\end{array}
\end{equation}
For any differential operator $n$ form $\hat{\bf \omega}^{n}$ on
$\hat{\cal D}_{2}$ the reflection $f$ induces the operator $n$ form
$\hat{f}^{*}\hat{\bf \omega}^{n}$ on $\hat{\cal D}_{1}$
\begin{equation}
\label{R427}
\left(\hat{f}^{*}\hat{\bf \omega}^{n}\right)
\left.\left(\hat{\bf A}_{1},\ldots,
\hat{\bf A}_{n}\right)\right|_{\Phi_{p}}=\hat{f}_{*}\hat{\bf \omega}^{n}
\left.\left(\hat{f}_{*}\hat{\bf A}_{1},\ldots,\hat{f}_{*}
\hat{\bf A}_{n}\right)\right|_{f\left(\Phi_{p}\right)}.
\end{equation}
If $\hat{\bf \omega'}^{1}=\alpha'_{i}d\,{\Phi'}^{\widehat{\imath}}$
then
$
\hat{f}^{*}\left(\alpha'_{i}d\,{\Phi'}^{\widehat{\imath}}\right)=
\alpha'_{i}\FFr{\partial {\Phi'}^{i}}{\partial {\Phi}^{j}}
d\,\Phi^{\widehat{\jmath}}.
$
This can be extended up to
$\hat{\bf \omega'}^{n} \rightarrow \hat{\bf \omega}^{n}$
\begin{equation}
\label{R429}
\hat{f}^{*}\left(
\S_{i_{1}<\cdots<i_{n}}
{T'}_{i_{1} \cdots i_{n}} d\,{\Phi'}^{\widehat{\imath_{1}}}\wedge\cdots\wedge
d\,{\Phi'}^{\widehat{\imath_{n}}}\right)
=\S_{\begin{array}{l}
{\scriptstyle i_{1}<\cdots<i_{n}} \\
{\scriptstyle j_{1}<\cdots<j_{n}}
\end{array}}
{T'}_{i_{1} \cdots i_{n}} 
\FFr{\partial {\Phi'}^{i_{1}}}{\partial {\Phi}^{j^{1}}}\cdots
\FFr{\partial {\Phi'}^{i_{n}}}{\partial {\Phi}^{j^{n}}}
d\,{\Phi}^{\widehat{\imath_{1}}}\wedge\cdots\wedge
d\,{\Phi}^{\widehat{\imath_{n}}},
\end{equation}
namely
$
\hat{f}^{*}\hat{\bf \omega'}^{n} = J_{\Phi}\hat{\bf \omega}^{n}=
\left( det\,d f\right)\hat{\bf \omega}^{n},
$
where $J_{\Phi}$ is the Jacobian of reflection $J_{\Phi}=\left\|
\FFr{\partial {\Phi'}^{i}}{\partial {\Phi}^{j}}\right\|$.
While
$$
\left(\hat{f}_{1}\circ\hat{f}_{2}\right)^{*}=
\hat{f}^{*}_{1}\circ\hat{f}^{*}_{2},\quad
\hat{f}^{*}\left(\hat{\bf \omega}_{1}\wedge
\hat{\bf \omega}_{2}\right)=
\hat{f}^{*}\left(\hat{\bf \omega}_{1}\right)\wedge
\hat{f}^{*}\left(\hat{\bf \omega}_{2}\right).
$$
We may consider the integration of operator $n$ form implying
$
\IIn_{\hat{\cal D}_{1}}\hat{f}^{*}\hat{\bf \omega}^{n} = 
\IIn_{\hat{\cal D}_{2}}\hat{\bf \omega}^{n}.
$
In general, let the $\hat{\cal D}_{1}$ is a limited convex
$n$ dimensional parallelepiped in $n$ dimensional operator space
$\hat{\bf R}^{n}$. One defines the $n$ dimensional $i$-th piece of 
integration path $\hat{\sigma}^{i}$ in $\hat{G}$ as $\hat{\sigma}^{i}=
\left(\hat{\cal D}_{i}, f_{i}, Or_{i}\right)$, where $\hat{\cal D}_{i}
\in \hat{\bf R}^{n}, \quad f_{i}:\hat{\cal D}_{i}\rightarrow \hat{G}$
and the $Or_{i}$ is an orientation of $\hat{\bf R}^{n}$. Then, the integral
over the operator $n$ form $\hat{\bf \omega}^{n}$ along the operator
$n$ dimensional chain $\hat{c}_{n}=\S m_{i}\hat{\sigma}^{i}$ may be
written
$$
\IIn_{\hat{c}_{n}}\hat{\bf \omega}^{n}=\S m_{i}
\IIn_{\hat{\sigma}^{i}}\hat{\bf \omega}^{n}=\S m_{i}
\IIn_{\hat{\cal D}_{i}}\hat{f}^{*}\hat{\bf \omega}^{n},
$$
where the $m_{i}$ is a multiple number. Taking into account the eq.(2.4.3),
the matrix element yields
\begin{equation}
\label{R434}
<\chi \| \IIn_{\hat{c}_{n}}\hat{\bf \omega}^{n}\|
\chi^{0} >
\rightarrow 
\left\{\S_{\lambda\mu=1}^{2}\right\}_{1}^{n}
\S_{r_{1}^{\lambda\mu},\ldots ,r_{n}^{\lambda\mu}=1}^{N_{\lambda\mu}}
\S m_{i}\bar{c}(r_{1}^{\lambda\mu},\ldots ,r_{n}^{\lambda\mu})
\IIn_{\hat{\cal D}_{i}}\hat{f}^{*}\hat{\bf \omega}^{n}
(r_{1}^{\lambda\mu},\ldots ,r_{n}^{\lambda\mu}).
\end{equation}
Next we employ the analog of exterior differentiation. We define 
the operator $(n+1)$ form $\hat{d}\,\hat{\bf \omega}^{n}$ on $(n+1)$
operator vectors $\hat{\bf A}_{1},\ldots,
\hat{\bf A}_{n+1}\in \hat{\bf T}_{\Phi_{p}}$ by considering 
diffeomorphic reflection $f$ of the neighbourhood of the point
$0$ in $\hat{\bf R}^{n}$ into neighbourhood of the point ${\bf \Phi}_{p}$
in $\hat{G}$. The prototypes of operator vectors
$\hat{\bf A}_{1},\ldots,
\hat{\bf A}_{n+1}\in \hat{\bf T}_{\Phi_{p}}\left(\hat{G}\right)$
at the operator differential of $f$ belong to tangent operator space
$\hat{\bf R}^{n}$ in $0$. Then, the prototypes are the operator vectors
$\hat{\bf \xi}_{1},\ldots,\hat{\bf \xi}_{n+1} \in \hat{\bf R}^{n}$. Let
$f$ reflects the parallelepiped $\hat{\bf \Pi}^{*},$ stretched over 
the $\hat{\bf \xi}_{1},\ldots,\hat{\bf \xi}_{n+1}$, into $(n+1)$
dimensional piece $\hat{\bf \Pi}$ on $\hat{G}$.
While the border of $n$ dimensional chain $\partial \hat{\bf \Pi}$ 
in $\hat{\bf R}^{n+1}$ defined as follows: the pieces $\hat{\sigma}^{i}$
of the chain $\partial \hat{\bf \Pi}$ are $n$ dimensional facets
$\partial \hat{\bf \Pi}_{i}$ of parallelepiped $\partial \hat{\bf \Pi}$
with the reflections embedding the facets into $\hat{\bf R}^{n+1}$:
$\quad f_{i}:\hat{\bf \Pi}_{i} \rightarrow
\hat{\bf R}^{n+1}$, and the
orientations $Or_{i}$ defined $\partial \hat{\bf \Pi}=\S\hat{\sigma}^{i},
\quad \hat{\sigma}^{i}=\left(\hat{\bf \Pi}_{i},f_{i},Or_{i}\right)$
Considering the curvilinear parallelepiped
$$
F\left(\hat{\bf A}_{1},\ldots,\hat{\bf A}_{n}\right)=
\IIn_{\partial \hat{\bf \Pi}}\hat{\bf \omega}^{n},
$$
one may state that the unique operator of $(n+1)$-form $\hat{\Omega}$
exists on $\hat{\bf T}_{\Phi_{p}}$, which is the principle $(n+1)$ linear
part in $0$ of integral over the border of 
$F\left(\hat{\bf A}_{1},\ldots,\hat{\bf A}_{n}\right)$, namely
\begin{equation}
\label{R436}
F\left(\varepsilon\hat{\bf A}_{1},\ldots,
\varepsilon\hat{\bf A}_{n}\right)=
\varepsilon^{n+1}\hat{\Omega}
\left(\hat{\bf A}_{1},\ldots,\hat{\bf A}_{n+1}\right)+
O\left(\varepsilon^{n+1}\right),
\end{equation}
where the $\hat{\Omega}$ is independent of the choice of coordinates
used in definition of $F$. 
The prove of it is similar to those of 
corresponding theorem of differential geometry [14].
If in local coordinates 
$
\hat{\bf \omega}^{n}=\S_{i_{1}<\cdots<i_{n}}
T_{i_{1} \cdots i_{n}} d\,\Phi^{\widehat{\imath_{1}}}\wedge\cdots\wedge
d\,\Phi^{\widehat{\imath_{n}}},
$
then
\begin{equation}
\label{R438}
\hat{\Omega}=\hat{d}\,\hat{\bf \omega}^{n}=
\S_{i_{1}<\cdots<i_{n}}
\hat{d}\,T_{i_{1} \cdots i_{n}} d\,\Phi^{\widehat{\imath_{1}}}
\wedge\cdots\wedge d\,\Phi^{\widehat{\imath_{n}}}.
\end{equation}
The operator of exterior differential $\hat{d}$ commutes with the
reflection 
$
f:\hat{G}\rightarrow \hat{G}
$
$$
\hat{d}\,\left(\hat{f}^{*}\hat{\bf \omega}^{n}\right)=
\hat{f}^{*}\left(\hat{d}\,\hat{\bf \omega}^{n}\right).
$$
So define the exterior differential by operator (n+1) form
\begin{equation}
\label{R440}
\hat{d}\hat{\omega}^{n}=\S_{\begin{array}{l}
{\scriptstyle i_{0}} \\
{\scriptstyle i_{1}<\ldots<i_{n}}
\end{array}}
\FFr{\partial T_{i_{1}\ldots i_{n}}}{\partial\Phi^{i_{0}}}
d\Phi^{\widehat{i_{0}}}\wedge
d\Phi^{\widehat{i_{1}}}\wedge\cdots\wedge d\Phi^{\widehat{i_{n}}}
=\S_{i_{1}<\ldots<i_{n}}(\hat{d}T_{i_{1}\ldots i_{n}})\wedge
d\Phi^{\widehat{i_{1}}}\wedge\cdots\wedge d\Phi^{\widehat{i_{n}}},
\end{equation}
then
\begin{equation}
\label{R441}
\begin{array}{l}
<\chi\parallel \hat{d}\hat{\omega}^{n} \parallel\chi^{0}>\rightarrow \\
\S_{i_{1}<\ldots<i_{n}}
\left\{\S_{\lambda,\mu=1}^{2}\right\}_{1}^{n}
\S_{r_{1}^{\lambda\mu},\ldots ,r_{n}^{\lambda\mu}=1}^{N_{\lambda\mu}}
\bar{c}(r_{1}^{\lambda\mu},\ldots r_{n}^{\lambda\mu})
{(dT(r_{1}^{\lambda\mu},\ldots ,r_{n}^{\lambda\mu})}_{i_{1}\ldots i_{n}})
\wedge
d\Phi^{i_{1}}_{r_{1}^{\lambda\mu}}\wedge\cdots\wedge 
{d\Phi}^{i_{n}}_{r_{n}^{\lambda\mu}}.
\end{array}
\end{equation}
We may draw a conclusion that the matrix element 
of any geometric object of operator 
manifold $\hat{G}$ yields corresponding geometric object of wave 
manifold $\cal {G}$.

\section{Primordial Structures and Link Establishing Processes}
\label{goyaks}
To facilitate the physical picture and provide sufficient background it
seems worth to bring few formal matters in concise form which one will 
have to know in order to understand the general structure of our approach
without undue hardship. Here we only outline briefly the relevant steps.
In the mean time we refer to [8] for more detailed justification of 
some of the procedures and complete exposition.\\
Before proceeding further, it is profitable to define 
the {\em pulsating gauge functions} 
and {\em fields} denoted by wiggles as follows:\\
1. An invariant with respect to the coordinate transformations 
function $\widetilde{W}(x)$ 
defined on the space $M$ ($x\in M$ ) is called the 
pulsating gauge function if it undergoes local gauge transformations
\begin{equation}
\label{R441}
\widetilde{W}'(x)=U(x)\widetilde{W}(x).
\end{equation}
Here $U(x)$ is the element of some simple Lie group $G$ the generators of 
which imply the algebra $[F^{a},F^{b}]=iC^{abc}F^{c}$, where $C^{abc}$
are wholly antisymmetric structure constants.\\
2. A smooth function $\widetilde{\Phi}(\widetilde{W}(x))$ belonged to
some representation of the group $G$, where the generators are presented
by the matrices $T^{a}$, is called the 
pulsating field if under the transformation eq.(3.0.1) it transforms
\begin{equation}
\label{R441}
\widetilde{\Phi}'\equiv
\widetilde{\Phi}(\widetilde{W}'(x))=U(x)\widetilde{\Phi}(\widetilde{W}(x)).
\end{equation}
Let $L_{0}(\Phi,\partial\Phi)$ is the invariant Lagrangian of free field
$\Phi$ defined on $M$. Then, a simple gauge invariant Lagrangian of the 
pulsating field $\widetilde{\Phi}$ can be written
\begin{equation}
\label{R441}
L={\widetilde{W}}^{+}\widetilde{W}L_{0}(\Phi,\partial\Phi),
\end{equation}
which reduces to
\begin{equation}
\label{R441}
L\equiv L\left(\widetilde{\Phi},\widetilde{D\Phi}\right)=
L_{0}\left(\widetilde{\Phi}(\widetilde{W}(x)),
D\widetilde{\Phi}(\widetilde{W}(x))\right).
\end{equation}
Here we have noticed that due to eq.(3.0.2) and eq.(3.0.1)  
$\widetilde{\Phi}(\widetilde{W})=\widetilde{W}\Phi$, and introduced the 
covariant derivative 
$\widetilde{D\Phi}\equiv D\widetilde{\Phi}(\widetilde{W})=
\widetilde{W}\partial\Phi.$
Whence
\begin{equation}
\label{R441}
D=\partial - igT^{a}W^{a}, \quad T^{a}W^{a}= -\FFr{i}{g}\partial
\ln \widetilde{W},
\quad D\widetilde{W}=\left(D\widetilde{W}\right)^{+}=0,
\end{equation}
where $W^{a}$ is the gauge field, g is the coupling constant.
Thus, the conventional matter fields interacting by gauge fields are
the pulsating fields. 

\setcounter{section}{3}
\subsection{The Regular Primordial Structures}
In [8] we have chosen a simple setting and considered the primordial 
structures designed to possess certain physical properties satisfying the 
stated general rules. These structures are the substance out of which the 
geometry and particles are made.
We distinguish $\eta$- and $u$-types
primordial structures involved in the linkage establishing processes 
occurring between the structures of different types.
Let us recall that the $\eta$-type structure may accept the linkage
only from $u$-type structure, which is described by the link
function 
${\ps1_{\eta} }(s)$
belonging to the ordinary class of functions of $C^{\infty}$ smoothness, 
where $s\equiv\eta={\e1_{\eta}}_{(\lambda\alpha)}
\eta^{(\lambda\alpha)},\quad (\lambda = \pm; \alpha = 1,2,3,$ see subsec.2.1),
$\eta$ is the link coordinate.
Respectively the $u$-type structure may accept the linkage only
from $\eta$-type structure described by the link function
$
{\ps1_{u}}(s)$ (u-channel), where
$s\equiv u=\e1_{u} u $.
We assume that $s$ is the pulsating gauge function associated with the Abelian local gauge group $U(1)$ and  $\Psi(s)$ is the
pulsating field (the wiggles are left implicit). Thus, 
under local gauge transformations 
$$
s'=e^{-i\alpha}s,\quad \partial\alpha\neq 0,
$$
the link function 
$\Psi(s)$ transforms 
$$\Psi(s')=e^{-i\alpha}\Psi(s),$$
and the Lagrangian eq.(3.0.3) is invariant under gauge 
transformations.
It includes the covariant derivative 
$D(s)=\partial +igb(s)$ and
gauge field $b(s)=\FFr{i}{g}\partial \ln s$ undergone gauge transformations
$
b(s')=b(s)+\FFr{1}{g}\partial\alpha.
$
Then  ${\ps1_{i}}(s)=s{\ps1_{i}}=
{\e1_{i}}_{(\lambda\alpha)}{\ps1_{i}}^{(\lambda\alpha)}$ ($i=\eta, u$), 
where the eq.(2.1.9) holds
\begin{equation}
\label {eq: R2.2}
{\ps1_{\eta} }^{(\pm\alpha)}(\eta,p_{\eta})=
\eta^{(\pm\alpha)}
{\ps1_{\eta} }^{\pm}(\eta,p_{\eta}),\quad 
{\ps1_{u} }^{(\pm\alpha)}(u,p_{u})=
u^{(\pm\alpha)}
{\ps1_{u} }^{\pm}(u,p_{u}),
\end{equation}
a bispinor ${\ps1_{i} }^{\pm}$ is the invariant state wave function of 
positive or negative frequencies, $p_{i}$ is the corresponding link momentum.
Thus, a primordial structure can be considered as
a fermion found in external gauge field 
$b(s). $\\
The simplest system made of two structures of different types becomes
stable only due to the stable linkage, namely 
\begin{equation}
\label {R72}
\left|\p1_{\eta}\right|={({\p1_{\eta}}^{(\lambda\alpha)},
{\p1_{\eta}}_{(\lambda\alpha)})}^{1/2}=
\left|\p1_{u}\right|={({\p1_{u}}^{(\lambda\alpha)},
{\p1_{u}}_{(\lambda\alpha)})}^{1/2}.
\end{equation}
Otherwise they are unstable.
There is not any restriction on the number of primordial structures
of both types getting into the link establishing processes simultaneously.
In the stable system the link stability condition must be
held for each linkage separately.\\
The persistent processes of creation and annihilation of the primordial 
structures occur in different states $s, s',s'',...$ The "creation"
of structure in the given state $(s)$ is due to its transition to 
this state from other states $(s',s'',...)$, while the "annihilation"
means a vice versa.
Satisfying eq.(3.1.2) the primordial structures from the given state 
as well as different states can establish a stable linkage.
Among the states $(s,s',s'',...)$ there is a lowest one ($s_{0}$),
in which all structures are regular. That is, they are in free (pure) state
and described by the plane wave functions
${\ps1_{\eta} }^{\pm }(\eta_{f},p_{\eta})$ or 
${\ps1_{u} }^{\pm }(u_{f},p_{u})$
defined respectively on flat manifolds
$\G1_{\eta}$ and $\G1_{u}$. The index (f) specifies the points of 
corresponding flat manifolds $\eta_{f}\in\G1_{\eta}$, $u_{f}\in
\G1_{u}$. For example, in accordance with subsec.2.2, the equation of 
regular structure $\Psi(s_{+})\quad (s=s_{+}+s_{-})$ reads
$$
\left[i\gamma_{f}(\partial +igb(s_{+})) -m\right]\Psi(s_{+})=0,
$$
the matrices $\gamma_{f}$ are given in eq.(3.3.3).
Whence the equation of plane wave function  
$\Psi^{+}_{p}$ of positive frequencies stems
$$
(i\gamma_{f} \partial - m)\Psi^{+}_{p}=0.
$$
The processes of creation and annihilation of regular structures in lowest 
state are described by the operator manifold formalism given in sec.2.

\subsection{The Distorted  Primordial Structures}
\label{Vec}
In all higher states the  primordial structures are distorted (interaction 
states) and described by distorted link functions defined on 
distorted manifolds $\widetilde{\G1_{\eta}}$ and $\widetilde{\G1_{u}}$.
The distortion 
$G\rightarrow \widetilde{G}$ 
with hidden Abelian local group
$G=U^{loc}(1)=SO^{loc}(2)$ and one dimensional trivial algebra
$\hat{g}=R^{1}$ is considered in [8,15].
It involves a drastic revision of a role of 
local internal symmetries in the concept of curved geometry.
Under the reflection of fields and their dynamics from Minkowski space 
to Riemannian a standard gauge principle of local internal symmetries is 
generalized. The gravitation gauge group is proposed, which is generated 
by hidden local internal symmetry.
This suggests an opportunity for the unification of all 
interactions on an equal footing.\\
Our scheme is implemented as follows:
Considering the principle bundle $p:E\rightarrow G$
the basis $e^{f}$ is transformed 
$
e=D\,e^{f},
$
under massless gauge distortion field $a_{f}$ associated with $U^{loc}(1)$.  
The matrix $D$ is in the form $D=C\otimes R$, where the distortion
transformations $O_{(\lambda\alpha)}=C^{\tau}_{(\lambda\alpha)}
O_{\tau}$ and
$\sigma_{(\lambda\alpha)}=R^{\beta}_{(\lambda\alpha)}
\sigma_{\beta}$ are defined. Here $C^{\tau}_{(\lambda\alpha)}=
\delta^{\tau}_{\lambda} + \kappa a_{(\lambda\alpha)}{}^{*}
\delta^{\tau}_{\lambda}$,
but $R$ is a matrix of the group $SO(3)$ of ordinary rotations of 
the planes involving two arbitrary basis vectors 
of the spaces $R^{3}_{\pm}$ around  the orthogonal third axes. 
The rotation angles are determined from the constraint imposed
upon distortion transformations that a sum of distorted parts
of corresponding basis vectors $O_{\lambda}$ and 
$\sigma_{\beta}$ should be zero at given $\lambda$
\begin{equation}
\label{R52}
<O_{(\lambda\alpha)},O_{\tau}>_{\tau \neq \lambda}+\frac{1}{2}
\varepsilon_{\alpha\beta\gamma}\frac{<\sigma_{(\lambda\beta)},\sigma_{\gamma}>}
{<\sigma_{(\lambda\beta)},\sigma_{\beta}>}=0,
\end{equation}
where $\varepsilon_{\alpha\beta\gamma}$ is an antisymmetric unit tensor.
Thereupon $\tan\theta_{(\lambda\alpha)}=-\kappa a_{(\lambda\alpha)}$,
where $\theta_{(\lambda\alpha)}$ is the particular rotation around the axis
$\sigma_{\alpha}$. Since the $R$
is independent of the sequence of rotation axes, then  it implies
the mean value $R=\displaystyle \frac{1}{6} \sum_{i \neq j \neq k}
R^{(ijk)}$, where $R^{(ijk)}$ the matrix of rotations occurring
in the given sequence $(ijk)$ $(i,j,k=1,2,3)$.
The field $a_{f}$ is due to the distortion of basis
pseudovector $O_{\lambda}$, while the distortion of $\sigma_{\alpha}$
follows from eq.(3.2.2).\\
Next we construct
the diffeomorphism $G\rightarrow \widetilde{G}$ and introduce the invariant 
action of the fields.
The passage from six dimensional curved manifold $\widetilde{G}$
to four dimensional Riemannian geometry $R^{4}$ is straightforward by
making use of reduction of three time components $e_{0\alpha}=
\displaystyle \frac{1}{\sqrt{2}}(e_{(+\alpha)}+e_{(-\alpha)})$ of basis 
sixvector $e_{(\lambda \alpha)}$ to the single one $e_{0}$ in the given
universal direction, which merely fixed a time 
coordinate. Actually, since Lagrangian of the fields defined on
$\widetilde{G}$ is a function of scalars, namely,
$A_{(\lambda \alpha)}B^{(\lambda \alpha)}=
A_{0 \alpha}B^{0 \alpha}+A_{\alpha}B^{\alpha}$, so taking into account that
$A_{0 \alpha}B^{0 \alpha}=A_{0 \alpha}<e^{0\alpha},e^{0\beta}>B_{0 \beta}
=A_{0}<e^{0},e^{0}>B_{0}=A_{0}B^{0}$, one readily may perform the
required passage.
In this case, instead of eq.(2.2.5), one has
\begin{equation}
\label{R923}
d\,\zeta^{2}=d\,\eta^{2}-d\,u^{2}=0,
\quad \left.d\,\eta^{2} \right|_{6\rightarrow 4} \equiv
d\,s^{2}=g_{\mu\nu}d\,x^{\mu}d\,x^{\nu}=d\,u^{2}=inv.
\end{equation}

\subsection{Reflection of the Fermi Fields}
\label{ref}
Within this approach we may consider the reflection
of the Fermi fields and their dynamics from the flat manifold $\G1_{u}$ 
to
distorted manifold $\widetilde{\G1_{u}}$, and vice versa. 
Thereat we construct a diffeomorphism
$u(u_{f}):\G1_{u}\rightarrow \widetilde{\G1_{u}}$, where
the holonomic functions 
$u(u_{f})$ satisfy defining
relation
\begin{equation}
\label {eq: RB.4}
e\psi=e^{f} + \chi^{f}({\bf B}_{f}).
\end{equation}
Here $e^{f}$ and $e$ are the basis vectors
on $\G1_{u}$ and $\widetilde{\G1_{u}}$.
The $\psi$ is taken to denote
$
\psi\equiv 
\FFr{\partial\,u}{\partial\,u_{f}}.
$
The covector
\begin{equation}
\label {eq: RB.7}
\chi^{f}_{(\tau\beta)}({\bf B}_{f})=
e_{(\lambda\alpha)}\chi^{(\lambda\alpha)}_{(\tau\beta)}=
-\FFr{1}{2}e_{(\lambda\alpha)}\IIn_{0}^{u_{f}}
({\pr_{u}}^{f}_{(\rho\gamma)}D^{(\lambda\alpha)}_{(\tau\beta)}-
{\pr_{u}}^{f}_{(\tau\beta)}D^{(\lambda\alpha)}_{(\rho\gamma)})
du^{(\rho\gamma)}_{f}
\end{equation}
realizes the coordinates $u$ 
by providing a criteria of integration and undegeneration [16,17]\footnote 
{I wish to thank S.P.Novikov for valuable discussion of this point}.
A Lagrangian $L(x)$ of fields $\Psi(u)$
may be obtained under the reflection from a Lagrangian $L_{f}(u_{f})$ of 
corresponding {\em shadow fields} $\Psi_{f}(u_{f})$ and vice versa.
The $\Psi_{f}(u_{f})$ is defined as the section of vector bundle 
associated with the primary gauge group G by reflection 
$\Psi_{f}:  \G1_{u}\rightarrow E$ that
$p\,\Psi_{f}(u_{f})=u_{f}$, where 
$u_{f} \in  \G1_{u}$ is a point of flat manifold  $\G1_{u}$ 
(specified by index $({}_{f})$). The $\Psi_{f}$ takes value in standard 
fiber $F_{u_{f}}$
upon $u_{f} : p^{-1}(U^{(f)})=U^{(f)}\times F_{u_{f}}$, where 
$U^{(f)}$ is a region of base of principle bundle 
upon which an expansion into direct product $p^{-1}(U^{(f)})=
U^{(f)}\times G$ is defined.  The fiber is Hilbert vector space 
on which a linear representation $U_{f}$ of the group G is given. 
Respectively
$\Psi(u)\subset F_{u}$, where $F_{u}$
is the fiber upon $u:p^{-1}(U)=U\times F_{u}$, $U$ is the region of base 
 $\widetilde{\G1_{u}}$. Thus, the reflection of bispinor fields may be written down
\begin{equation}
\label {eq: RB.1}
\begin{array}{l}
\Psi(u)=R({\bf B}_{f})\Psi_{f}(u_{f}), \quad
\bar{\Psi}(u)=\bar{\Psi}_{f}(u_{f})\widetilde{R}^{+}({\bf B}_{f}),\\\\
g(u)\nabla\Psi(u)=S(B_{f})R({\bf B}_{f})
\gamma_{f}D\Psi_{f}(u_{f}),\\ \\
\left( \nabla\bar{\Psi}(u)\right)g(u)
=S(B_{f})\left( D\bar{\Psi}_{f}(u_{f})\right)
\gamma_{f}\widetilde{R}^{+}({\bf B}_{f}).
\end{array}
\end{equation}
Reviewing the notation
${\bf B}(u_{f})=T^{a}{\bf B}^{a}(u_{f})$
is the gauge field of distortion
with the values in Lie algebra of group G, $R({\bf B}_{f})$ is the reflection 
matrix (see eq.(3.3.6)),
$\widetilde{R}=\gamma^{0}R\gamma^{0},$
$D=\partial^{f}- ig {\bf B},
\quad g^{(\lambda\alpha)}(\theta)=V^{(\lambda\alpha)}_{(i,l)}
(\theta){\gamma}^{(i,l)}_{f}$, 
$V^{(\lambda\alpha)}_{(i,l)}(\theta)$ are congruence parameters of curves
(Latin indices refer to tetrad components).
The matrices
${\gamma}^{(\pm\alpha)}_{f}=\FFr{1}{\sqrt{2}}(\gamma^{0}\sigma^{\alpha}
\pm \gamma^{\alpha})$, 
$\gamma^{0},\gamma^{\alpha}$ are Dirac matrices. 
$\nabla$ is 
covariant derivative defined on  $\widetilde{\G1_{u}}$:
$\nabla={\pr_{u}}+
\Gamma$, where
the connection
$\Gamma(\theta)$ in terms of Ricci 
rotation coefficients reads
$
\Gamma_{(\lambda\alpha)}(\theta)=\FFr{1}{4}
\Delta_{(\lambda\alpha)(i,l)(m,p)}\gamma^{(i,l)}_{f}\gamma^{(m,p)}_{f},
\quad
\bar{\Gamma}_{(\lambda\alpha)}(\theta)=\FFr{1}{4}
\Delta_{(\lambda\alpha)(i,l)(m,p)}\gamma^{(m,p)}_{f}\gamma^{(i,l)}_{f}.
$\\
According to the general gauge principle [8,15], the physical system of
the fields $\Psi(u)$ is required to be invariant under the finite 
local gauge transformations 
\begin{equation}
\label {eq: RB.8}
\begin{array}{l}
\Psi'(u)=U_{R}\Psi(u),\\\\
\left(
g(u) \nabla \Psi(u)\right)'=
U_{R}\left(g(u)\nabla \Psi(u)
\right), \quad U_{R}=R({\bf B}'_{f})U_{f}R^{-1}({\bf B}_{f}),
\end{array}
\end{equation}
of the Lie group of gravitation $G_{R} (\ni  U_{R})$ generated by G, 
where the gauge field
${\bf B}_{f}(u_{f})$ is transformed under 
G in standard form.
The physical meaning of the general principle is as follows: one has 
conventional G-gauge theory on flat manifold in terms of curviliniear 
coordinates if curvature tensor is zero, to which the zero vector 
eq.(3.3.2) is corresponded.
Otherwise it yields the gravitation interaction. \\
Out of a set of arbitrary curvilinear coordinates in $\widetilde{\G1_{u}}$ 
the
real curvilinear coordinates may be distinguished, which satisfy
eq.(3.3.1) under all possible Lorentz and gauge transformations. There is
a single~-valued conformity between corresponding tensors with various
suffixes on $\widetilde{\G1_{u}}$ and $ \G1_{u}$. While, each 
index transformation is incorporated with function $\psi$. 
The transformation of real curvilinear coordinates
$u \rightarrow u'$ is due to some Lorentz $(\Lambda)$ and gauge
$(B_{f}\rightarrow B'_{f})$ transformations
\begin{equation}
\label {R213}
\frac{\partial u'}{\partial u}=\psi(B'_{f})
\psi(B_{f})\Lambda.
\end{equation}
There would then exist preferred systems and group of transformations of
real curvilinear coordinates in  $\widetilde{\G1_{u}}$. The wider group of 
transformations of arbitrary curvilinear coordinates in  
$\widetilde{\G1_{u}}$ 
would then be of no consequence for the field dynamics.
A straightforward calculation gives
the reflection matrix 
\begin{equation}
\label{eq: R2.8}
R(u,u_{f})=R_{f}(u_{f})R_{g}(u)=\exp\left[ 
-i\Theta_{f}(u_{f})-
\Theta_{g}(u)\right],
\end{equation}
where
\begin{equation}
\label{eq: R2.9}
\Theta_{f}(u_{f})=
g\IIn_{0}^{u_{f}}{\bf B}(u_{f})du_{f},
\quad 
\Theta_{g}(u)=\frac{1}{2}\IIn_{0}^{u} R^{+}_{f}\left\{ 
g\Gamma R_{f}, g \,du\right\}.
\end{equation}
Then
\begin{equation}
\label {eq: RB.9}
S(B_{f})=\frac{1}{8K}\psi\left\{\widetilde{R}^{+}g\,R,
\gamma^{0}\right\}=inv,
\end{equation}
where
\begin{equation}
\label {eq: RB.9}
K =\widetilde{R}^{+}R=\widetilde{R}^{+}_{g}R_{g}
= 1.
\end{equation}
and
\begin{equation}
\label {eq: RB.13}
\widetilde{U}_{R}^{+}U_{R}=\gamma^{0}U_{R}^{+}\gamma^{0}U_{R}=1.
\end{equation}
The Lagrangian of shadow Fermi field may be written
\begin{equation}
\label {eq: RB.15}
\begin{array}{l}
L_{f}(u_{f})=J_{\psi} L(u)=\\
J_{\psi}\left\{S(B_{f})
\displaystyle {\frac{i}{2}}\left[ 
\bar{\Psi}_{f}(u_{f})\gamma_{0}D
\Psi_{f}(u_{f})-\right.\right. 
\left.\left.
\left(D\bar{\Psi}_{f}(u_{f})\right)\gamma_{0}
\Psi_{f}(u_{f})\right]-
m\bar{\Psi}_{f}(u_{f})\Psi_{f}(u_{f})\right\}.
\end{array}
\end{equation}
provided by
$
J_{\psi}= \| \psi\|\,\sqrt{-g} \equiv\left(1+2\|
<e^{f},\chi^{f}>\|
+ \| <\chi^{f},\chi^{f}>\|\right)^{1/2}.
$
The Lagrangian $L(u)$ of the field $\Psi(u)$ reads
\begin{equation}
\label {eq: RB.16}
\sqrt{-g}L(u)=\FFr{\sqrt{-g}}{2} \left\{-i \bar{\Psi}(u)
g({\pr_{u}}-\Gamma)\Psi(u)+ i\bar{\Psi}(u)
({\lpr_{u}}-\bar{\Gamma})g\bar{\Psi}(u)+
2m\bar{\Psi}(u)\Psi(u)\right\},
\end{equation}
yielding the field equations
\begin{equation}
\label {eq: RB.18}
\left[ i g ({\pr_{u}}-\Gamma )- m \right]\Psi(u) =0,\quad
\bar{\Psi}(u)\left[
i({\lpr_{u}}-\bar{\Gamma}) g - m \right] =0.
\end{equation}
The solution enables to write down the relation between the  
wave functions of distorted and regular structures [8] 
\begin{equation}
\label{eq: R2.4}
{\ps1_{u}}^{\lambda}(\theta_{+k})=
f_{(+)}(\theta_{+k}){\ps1_{u}}^{\lambda}, 
\quad
{\ps1_{u}}_{\lambda}(\theta_{-k})=
{\ps1_{u}}_{\lambda}f_{(-)}(\theta_{-k}).
\end{equation}
The ${\ps1_{u}}^{\lambda}({\ps1_{u}}_{\lambda})$ is the plane wave function 
of regular ordinary structure (antistructure) and
\begin{equation}
\label{eq: R2.5}
f_{(+)}(\theta_{+k})=e^{\chi_{R}(\theta_{+k})-
i\chi_{J}(\theta_{+k})}, \quad
f_{(-)}(\theta_{-k})=\left.f_{(+)}^{*}(\theta_{+k}) 
\right|_{\theta_{+k}=\theta_{-k}},
\end{equation}
where the $\chi_{R}$ and $\chi_{J}$ are given in Appendix.\\
Next we shall admit that 
the $\eta$-type (fundamental) regular structure can not directly form
a stable system with the regular $u$-type (ordinary) structures.
Instead of it the $\eta$-type regular structure 
forms a stable system with the infinite number of distorted ordinary 
structures, where the link stability 
condition held for each linkage separately. Such structures take part 
in realization of flat manifold $G$ (subsec.2.2). 
The laws regarding to this apply in use of 
functions of distorted ordinary structures 
\begin{equation}
\label{eq:R2.6}
{\ps1_{u}}^{(\lambda\alpha)}(\theta_{+})=u^{(\lambda\alpha)}
{\ps1_{u}}^{\lambda}(\theta_{+}),\quad
{{\ps1_{u}}}_{(\lambda\alpha)}(\theta_{-})=u_{(\lambda\alpha)}
{\ps1_{u}}_{\lambda}(\theta_{-}),
\end{equation}
where $u \in \widetilde{\G1_{u}}$.
We employ the wave packets constructed by superposition of these functions 
furnished by generalized operators of creation and annihilation as the 
expansion coefficients
\begin{equation}
\label{eq: R2.11}
\begin{array}{l}
{\hps_{u}}(\theta_{+})=\S_{\pm s}\IIn\frac{d^{3}p_{u}}{{(2\pi)}^{3/2}}
\left( {\hgam_{u}}_{(+\alpha)}^{k}{\ps1_{u}}^{(+\alpha)}(\theta_{+k})+
{\hgam_{u}}_{(-\alpha)}^{k}{\ps1_{u}}^{(-\alpha)}(\theta_{+k})\right), 
\\ 
{\bar{\hps_{u}}}(\theta_{-})=\S_{\pm s}\IIn\frac{d^{3}
p_{u}}{{(2\pi)}^{3/2}}
\left( {\hgam_{u}}^{(+\alpha)}_{k}{\ps1_{u}}_{(+\alpha)}(\theta_{-k})+
{\hgam_{u}}^{(-\alpha)}_{k}{\ps1_{u}}_{(-\alpha)}(\theta_{-k})\right),
\end{array}
\end{equation}
where as usual the summation is extended over all dummy indices.
The matrix element of anticommutator of generalized 
expansion coefficients reads 
\begin{equation}
\label{eq: R2.12}
<\chi_{-}\mid \{ {\hgam_{u}}^{(+\alpha)}_{k}(p,s),
{\hgam_{u}}_{(+\beta)}^{k'}(p',s') \}\mid\chi_{-}>= 
-\delta_{ss'}\delta_{\alpha\beta}\delta_{kk'}\delta^{3}(\vec{p}-\vec{p'}).
\end{equation}
The wave packets eq.(3.3.17) yield the causal Green's function
${\G1_{u}}^{\theta}_{F}(\theta_{+} - \theta_{-})$ of distorted 
ordinary structure.
Geometry realization requirement (eq.(2.2.4)) now must be satisfied for 
each ordinary structure in terms of
\begin{equation}
\label{eq: R2.13}
{\G1_{u}}^{\theta}_{F}(0)=\Lm_{\theta_{+} \rightarrow \theta_{-}}
{\G1_{u}}^{\theta}_{F}(\theta_{+} - \theta_{-})
={\G1_{\eta}}_{F}(0)= \Lm_{\eta'_{f} \rightarrow \eta_{f}}
{\G1_{\eta}}_{F}(\eta'_{f}-\eta_{f}).
\end{equation}
They are valid if following relations hold for
each distorted ordinary structure:
\begin{equation}
\label{eq: R2.14}
\S_{k}{\ps1_{u}}(\theta_{+k})
{\bar{\ps1_{u}}}(\theta_{-k})
=\S_{k}{\ps1_{u}}'(\theta'_{+k})
{\bar{\ps1_{u}}}'(\theta'_{-k})=
\cdots = inv. 
\end{equation}
Namely, 
the distorted ordinary structures only in permissible
combinations realize the geometry in a stable system.
Below, in schematic manner we exploit the background of the
Colour Confinement and Gauge principles.

\subsection{Quarks and Colour Confinement}
\label{quark}
We may think of the function
${\ps1_{u}}^{\lambda}(\theta_{+k})$ at fixed $(k)$ as being 
$u$-component of bispinor field of quark ${q}_{k}$, and of
${\bar{\ps1_{u}}}_{\lambda}(\theta_{-k})$ - an $u$-component of 
conjugated bispinor field of antiquark ${\bar{q}}_{k}$.
The index $(k)$ refers to colour degree of freedom in the 
case of rotations through the angles  $\theta_{+k}$ and anticolour 
degree of freedom in the case of $\theta_{-k}$.
The $\eta$-components of quark fields are plane waves.
In cases of local and global rotations we respectively distinguish 
two types of quarks: local ${q}_{k}$ and global ${q}_{k}^{c}$,
which will be in use  in the next Part II as the local and global subquark
fields.
Thus, the quark is a fermion with the half integer spin and certain 
colour degree of freedom.
There are exactly three colours.
The rotation through the angle $\theta_{+k}$ yields
a total quark field defined on the flat manifold $G=\G1_{\eta}\oplus\G1_{u}$
\begin{equation}
\label{eq: R2.15}
{q}_{k}(\theta)=\Psi(\theta_{+k})={\ps1_{\eta}}^{0}
\ps1_{u}(\theta_{+k})
\end{equation}
where ${\ps1_{\eta}}^{0}$ is a plane wave defined on $\G1_{\eta}$.
According to eq.(3.3.14), one gets
\begin{equation}
\label{eq: R2.16}
{q}_{k}(\theta)={\ps1_{\eta}}^{0}{\q1_{u}}_{k}(\theta)=
{\q1_{\eta}}_{k}(\theta){\ps1_{u}}^{0},
\quad {\q1_{\eta}}_{k}(\theta)\equiv f_{(+)}(\theta_{+k}){\ps1_{\eta}}^{0}, 
\end{equation}
where ${\ps1_{u}}^{0}$ is a plane wave, ${\q1_{u}}_{k}(\theta)$ and
${\q1_{\eta}}_{k}(\theta)$ may be considered as the quark fields with the 
same quantum numbers 
defined respectively on flat manifolds $\G1_{u}$ and  $\G1_{\eta}$. 
By making use of the rules stated in subsec.2.1
one may readily return to Minkowski space
$\G1_{\eta}\rightarrow M^{4}$. In the sequel, a conventional quark 
field defined on $M^{4}$ will be ensued
${\q1_{\eta}}_{k}(\theta) \rightarrow q_{k}(x)$, $x \in M^{4}$.
Due to eq.(3.3.20) and eq.(3.4.1) they imply
\begin{equation}
\label{eq: R2.17}
\S_{k}{q}_{kp}{\bar{q}}_{kp} =
\S_{k}{q'}_{kp}{\bar{q'}}_{kp}=
\cdots = inv,
\end{equation}
and
\begin{equation}
\label{eq:R2.18}
\S_{k}f_{(+)}(\theta_{+k})f_{(-)}(\theta_{-k} )
=\S_{k}f'_{(+)}(\theta'_{+k})f'_{(-)}(\theta'_{-k} )=
\cdots =inv.
\end{equation}
The eq.(3.4.3) utilizes the idea of Colour (Quark) Confinement 
principle: the quarks emerge in the geometry only in special combinations
of {colour singlets}.
Only two colour singlets are available (see below)
\begin{equation}
\label{eq: R2.20}
(q\bar{q})=\FFr{1}{\sqrt{3}}\delta_{kk'}{\hat{q}}_{k}{\bar{\hat{q}}}_{k'}=
inv, 
\quad
(qqq)=\FFr{1}{\sqrt{6}}\varepsilon_{klm}{\hat{q}}_{k}{\hat{q}}_{l}
{\hat{q}}_{m}=inv.
\end{equation}

\subsection {Gauge Principle; Internal Symmetries}
\label {gauge}
Following [8], the principle of identity holds for ordinary
regular structures, namely each regular structure in the lowest state
can be regarded as a result of transition from an
arbitrary state, in which they assumed to be distorted. 
This is stated below
\begin{equation}
\label{eq: R2.22}{\ps1_{u}}_{\lambda}=f^{-1}_{(+)}(\theta_{+k})
{\ps1_{u}}_{\lambda}(\theta_{+k})=
f^{-1}_{(+)}(\theta'_{+l})
{\ps1'_{u}}_{\lambda}(\theta'_{+l})=
\cdots. 
\end{equation}
Hence, the following transformations may be implemented upon
distorted ordinary structures
\begin{equation}
\label{eq: R2.21}
\begin{array}{l}
{\ps1_{u}}'^{\lambda}(\theta'_{+l})=
f^{(+)}_{lk}{\ps1_{u}}^{\lambda}(\theta_{+k})=
f(\theta'_{+l},\theta_{+k}){\ps1_{u}}^{\lambda}(\theta_{+k}),\\ 
{\ps1_{u}}'_{\lambda}(\theta'_{-l})=
{\ps1_{u}}_{\lambda}(\theta_{-k})f^{(-)}_{kl}=
\left.{\ps1_{u}}_{\lambda}(\theta_{-k})f^{*}(\theta'_{-l},\theta_{-k})
\right|_{\begin{array}{l}
\theta'_{-l}=\theta'_{+l}\\
\theta_{-k}=\theta_{+k},
\end{array}}
\end{array}
\end{equation}
provided
\begin{equation}
\label{eq: R2.22}
f^{(+)}_{lk}=\exp \{ \chi^{R}_{lk}-i\chi^{J}_{lk} \}, 
\quad \left. f^{(-)}_{kl}={(f^{(+)}_{lk})}^{*}
\right|_{\begin{array}{l}
\theta'_{-l}=\theta'_{+l}\\
\theta_{-k}=\theta_{+k},
\end{array}},
\end{equation} 
\begin{equation}
\label{eq: R2.22}
\chi^{R}_{lk}=\chi_{R}(\theta'_{+l})-\chi_{R}(\theta_{+k}),
\quad
\chi^{J}_{lk}=\chi_{J}(\theta'_{+l})-\chi_{J}(\theta_{+k}).
\end{equation}
The transformation functions are the operators in the space 
of internal degrees of freedom labeled by $(\pm k)$ corresponding to 
distortion rotations around the axes $(\pm k)$ by the
angles  $\theta_{\pm k}$. 
We make proposition that the distortion rotations are incompatible, namely the
transformation operators $f^{(\pm)}_{lk}$ obey the incompatibility relations
 \begin{equation}
\label{eq: R2.25}
\begin{array}{l}
f^{(+)}_{lk}f^{(+)}_{cd}-f^{(+)}_{ld}f^{(+)}_{ck}=\|f^{(+)}\|
\varepsilon_{lcm}\varepsilon_{kdn}f^{(-)}_{nm},\\ 
f^{(-)}_{kl}f^{(-)}_{dc}-f^{(-)}_{dl}f^{(-)}_{kc}=\|f^{(-)}\|
\varepsilon_{lcm}\varepsilon_{kdn}f^{(+)}_{mn},
\end{array}
\end{equation}
where $l,k,c,d,m,n=1,2,3$.
The relations eq.(3.5.5) hold in general for
both local and global rotations.
Making use of eq.(3.4.1), eq.(3.4.2) and eq.(3.5.2), one gets the transformations 
implemented upon the quark field, which in matrix notation take the form
$
q'(\zeta)=U(\theta(\zeta))q(\zeta), 
\quad
\bar{q'}(\zeta)=\bar{q}(\zeta)U^{+}(\theta(\zeta)),
$
where $q = \{ {q}_{k} \}, \quad U(\theta)=\{ f^{(+)}_{lk}  \}$. 
Due to the incompatibility commutation relations
(3.5.5), the transformation matrices $\{ U \}$  generate
the unitary group of internal symmetries $U(1), SU(2),$ 
$SU(3)$. 
As far as distorted ordinary structures have took
participation in the realization of geometry $G$ instead of
regular ones, stated somewhat differently
the principle of identity of regular structures 
directly leads to the equivalent one:
an action integral of any physical system must be invariant under
arbitrary transformations eq.(3.5.2) (the Gauge principle).
Below we discuss different possible models.\\
1. In the simple case of one dimensional local
transformations, through the local angles $\theta_{+1}(\zeta)$ and
$\theta_{-1}(\zeta)$
$
f^{(+)}=\left( \matrix{
f^{(+)}_{11}  & 0  & 0 \cr 
0             & 1  & 0 \cr
0             & 0  & 1\cr
}\right), \quad
f^{(-)}={(f^{(+)})}^{+}.
$
The incompatibility relations (3.5.5) 
reduce to identity $f^{(+)}_{11}=\|f^{(+)}\|.$
If
$\chi_{R}(\theta_{+1})=\chi_{R}(\theta_{-1})$,
and transformations
\begin{equation}
\label{eq: R2.28}
f^{(+)}_{11}=U(\theta)=f(\theta_{+1}(\zeta),\theta_{-1}(\zeta))=
\exp \{ -i\chi^{(+)}_{J}(\theta_{+1})+i\chi^{(-)}_{J}(\theta_{-1}) \}.
\end{equation}
generate a commutative
Abelian unitary local group of electromagnetic interactions  
realized as the Lie group $U^{loc}(1)= SO^{loc}(2)$ with
one dimensional trivial algebra $\hat{g}_{1}=R^{1}$:
$U(\theta)=e^{-i\theta}$,
where
$\theta \equiv \chi^{(+)}_{J}(\theta_{+1})-\chi^{(-)}_{J}(\theta_{-1})$.
The strength of interaction is specified by a single coupling $Q$ of 
electric charge. 
The invariance under
the local group $U^{loc}(1)$ leads to electromagnetic field, the massless
quanta of which - photons are {\em electrically neutral}, just because of the
condition eq.(3.4.4):
\begin{equation}
\label{eq: R2.29}
f(\theta_{+1},\theta_{-1})=f(\theta'_{+1},\theta'_{-1})=
\cdots inv.
\end{equation}
2. Next we consider a particular case of two dimensional local transformations
through the angles $\theta_{\pm m}(\zeta)$ around two axes $(m=1,2)$.
The matrix function of transformation is written down
$
f^{(+)}=\left( \matrix{
f^{(+)}_{11}  & f^{(+)}_{12}  & 0 \cr 
f^{(+)}_{21}    & f^{(+)}_{22}  & 0 \cr
0               & 0             & 1\cr
}\right), \quad
f^{(-)}={(f^{(+)})}^{+}.
$
The incompatibility relations (3.5.5) give rise to nontrivial conditions
\begin{equation}
\label{eq: R2.31}
\begin{array}{ll}
f^{(+)}_{11}=\left\| f^{(+)}\right\| {(f^{(+)}_{22})}^{*}, \qquad
f^{(+)}_{21}=-\left\| f^{(+)}\right\| {(f^{(+)}_{12})}^{*},\\ 
f^{(+)}_{12}=-\left\| f^{(+)}\right\| {(f^{(+)}_{21})}^{*},\quad
f^{(+)}_{22}=\left\| f^{(+)}\right\| {(f^{(+)}_{11})}^{*},
\end{array} 
\end{equation}
Hence $\left\| f^{(+)}\right\|=1.$ One readily infers the matrix 
$U(\theta)$ of gauge transformations of collection of fundamental fields
$
U=e^{-i\vec{T}\vec{\theta}}=
\left( \matrix{
f^{(+)}_{11}  & f^{(+)}_{12}\cr
f^{(+)}_{21}  & f^{(+)}_{22}\cr 
} \right),
$
where $T_{i} \quad(i=1,2,3)$ are the generators
of the group $SU(2)$.
The fields will come in multiplets forming a 
basis for representations of the isospin group $SU(2)$. Meanwhile
\begin{equation}
\label{eq: R2.33}
\begin{array}{l}
\FFr{\theta_{1}}{\theta}=\FFr{e^{\chi^{R}_{12}}\sin\chi^{J}_{12}}
{\sqrt{1-e^{2\chi^{R}_{11}}{\cos}^{2}\chi^{J}_{11}}}, 
\quad
\FFr{\theta_{2}}{\theta}=-\FFr{e^{\chi^{R}_{12}}\cos\chi^{J}_{12}}
{\sqrt{1-e^{2\chi^{R}_{11}}{\cos}^{2}\chi^{J}_{11}}},\quad
\FFr{\theta_{3}}{\theta}=\FFr{e^{\chi^{R}_{11}}\sin\chi^{J}_{11}}
{\sqrt{1-e^{2\chi^{R}_{11}}{\cos}^{2}\chi^{J}_{11}}}, \\ 
\theta = \mid\vec{\theta}\mid 
= 2\arccos \left(e^{\chi^{R}_{11}}\cos\chi^{J}_{11}\right),
\quad e^{\chi^{R}_{11}}\leq 1,
\end{array}
\end{equation}
provided
\begin{equation}
\label{eq:R2.34}
\chi^{R}_{11}=\chi^{R}_{22}, \quad\chi^{R}_{12}=\chi^{R}_{21}
\quad
\chi^{J}_{11}+\chi^{J}_{22}=0, \quad \chi^{J}_{21}+\chi^{J}_{12}=\pi,
\quad
\chi^{R}_{12}=\FFr{1}{2}\ln \left(1-e^{2\chi^{R}_{11}}\right).
\end{equation}
That is, three functions $\chi^{R}_{11}, \chi^{J}_{11}$ and $\chi^{J}_{12}$
or the angles $\theta'_{+1},\theta_{+1}$ and $\theta_{+2}$ are parameters
of the group $SU^{loc}(2)$ 
\begin{equation}
\label{eq; R2.35}
\chi^{R}_{11}=\chi_{R}(\theta'_{+1})-\chi_{R}(\theta_{+1}),
\quad
\chi^{J}_{11}=\chi_{J}(\theta'_{+1})-\chi_{J}(\theta_{+1}),
\quad
\chi^{J}_{12}=\chi_{J}(\theta'_{+1})-\chi_{J}(\theta_{+2}).
\end{equation}
3. In a case of gauge transformations occurred around all three
axes $(l,k=1,2,3)$:
$
f^{(+)}=\left( \matrix{
f^{(+)}_{11}  & f^{(+)}_{12}  & f^{(+)}_{13} \cr 
f^{(+)}_{21}  & f^{(+)}_{22}  & f^{(+)}_{23} \cr
f^{(+)}_{31}  & f^{(+)}_{32}  & f^{(+)}_{33}\cr
} \right), \quad
f^{(-)}={(f^{(+)})}^{+},
$
the incompatibility relations (3.5.5) yield the unitary condition
$U^{-1}=U^{+}, \,
f^{(+)}\equiv U$, and also $\left\| U \right\| =1$. Then
$U(\theta)=e^{-\frac{i}{2}\vec{\lambda}\vec{\theta}}$,
where $\FFr{\lambda_{i}}{2} \quad(i=1,\ldots,8)$ are the matrix
representation of generators of the group $SU(3)$.
Right through  differentiation one infers 
$\vec{\lambda}\vec{d\theta}=2iU^{+}dU,$
or
$
\vec{\theta}=-\IIn Im\left(tr \left(\vec{\lambda}
\left( f^{(-)}df^{(+)} \right) 
\right)\right),
$
provided
$Re\left(tr \left(\vec{\lambda}\left( f^{(-)}df^{(+)}\right) \right)
\right)\equiv 0.$
At the infinitesimal transformations $\theta_{i} \ll 1$, we get
\begin{equation}
\label{eq: R2.40}
\begin{array}{ll}
\theta_{1}\approx 2e^{\chi^{R}_{12}}\sin\chi^{J}_{12},\quad   
\theta_{3}\approx \sin\chi^{J}_{33}+2\sin\chi^{J}_{11},\quad       
\theta_{5}\approx 2(1- e^{\chi^{R}_{13}}\cos\chi^{J}_{13}),\\
\theta_{2}\approx 2(1- e^{\chi^{R}_{12}}\cos\chi^{J}_{12}),\quad  
\theta_{4}\approx 2e^{\chi^{R}_{13}}\sin\chi^{J}_{13},\quad       
\theta_{6}\approx 2e^{\chi^{R}_{23}}\sin\chi^{J}_{23},\\ 
\theta_{7}\approx 2(1- e^{\chi^{R}_{23}}\cos\chi^{J}_{23}),\quad
\theta_{8}\approx -\sqrt{3}\sin\chi^{J}_{33},
\end{array}
\end{equation}
provided
\begin{equation}
\label{eq:R2.41}
\chi^{R}_{ll}\approx 0,
\quad
\chi^{R}_{lk}\approx \chi^{R}_{kl}, \quad 
\chi^{J}_{lk}\approx \chi^{J}_{kl}, \quad (l \neq k)\quad
\sin\chi^{J}_{11}+\sin\chi^{J}_{22}+\sin\chi^{J}_{33}\approx 0.
\end{equation}
The internal symmetry group
$SU^{loc}_{C}(3)$ enables to introduce a gauge theory in colour space,
with the colour charges as exactly conserved quantities. The local colour
transformations are implemented on the coloured quarks right through a
$SU^{loc}_{C}(3)$ rotation matrix U in the fundamental 
representation. 

\section {Operator Multimanifold $\hat{G}_{N}$}
\label {Oper}
\subsection{Operator Vector and Covector Fields}
\label{ov}
The formalism of operator manifold $\hat{G}=
\hG_{\eta}\oplus\hG_{u}$ is built up by assuming 
an existence only of ordinary primordial structures of one sort 
(one u-channel).
Being confronted by our major
goal to develop the microscopic approach to field theory
based on multiworld geometry, 
henceforth instead of one sort of ordinary structures we are 
going to deal with different species of ordinary structures. That is,
before we enlarge the previous model we must make an additional
assumption concerning 
an existence of infinite number of ${}^{i}u$-type ordinary
structures of different species $i=1,2,\ldots ,N $ (multi-u channel).
These structures will be specified by the superscript
to the left. 
This hypothesis,
as it will be seen in the Part II, leads to 
the progress of understanding of the properties of particles.
At the very outset we consider the processes
of creation and annihilation of regular structures of $\eta$- and 
${}^{i}u$-types in the lowest state ($s_{0}$).
The general rules stated in subsec 2.1 regarding to this change apply 
a substitution of operator basis pseudo vectors and covectors by a new
ones supplied with additional superscript $(i)$ to the left referring
to different species $(i=1,2,\ldots, N)$
\begin{equation}
\label {eq: R3.7}
{}^{i}\hat{O}^{r_{1}r_{2}}_{\lambda,\mu}=
{}^{i}\hat{O}^{r_{1}}_{\lambda}\otimes
{}^{i}\hat{O}^{r_{2}}_{\mu}\equiv
{}^{i}\hat{O}^{r}_{\lambda,\mu}={}^{i}O^{r}_{\lambda,\mu}
(\alpha_{\lambda}\otimes \alpha_{\mu}),
\end{equation}
provided $ r=(r_{1},r_{2})$ and
$$
\begin{array}{l}
{}^{i}O^{r}_{1,1}=\FFr{1}{\sqrt{2}}(\nu_{i}{\O1_{\eta}}_{+}^{r}+
{}^{i}{\O1_{u}}_{+}^{r}), \quad
{}^{i}O^{r}_{2,1}=\FFr{1}{\sqrt{2}}(\nu_{i}{\O1_{\eta}}_{+}^{r}-
{}^{i}{\O1_{u}}_{+}^{r}), \\ 
{}^{i}O^{r}_{1,2}=\FFr{1}{\sqrt{2}}(\nu_{i}{\O1_{\eta}}_{-}^{r}+
{}^{i}{\O1_{u}}_{-}^{r}), \quad
{}^{i}O^{r}_{2,2}=\FFr{1}{\sqrt{2}}(\nu_{i}{\O1_{\eta}}_{-}^{r}-
{}^{i}{\O1_{u}}_{-}^{r}),
\end{array}
$$
where 
\begin{equation}
\label {eq: R3.9}
<\nu_{i},\nu_{j}>=\delta_{ij},\quad
<{}^{i}{\O1_{u}}_{\lambda}^{r},{}^{i}{\O1_{u}}_{\tau}^{r'}>=
-\delta_{ij}\delta_{rr'}{}^{*}\delta_{\lambda\tau},\quad
<{\O1_{\eta}}_{\lambda}^{r},{}^{i}{\O1_{u}}_{\tau}^{r'}>=0.
\end{equation}
In analogy with subsec.2.1 we consider the operator
$
{}^{i}{\hat{\gamma}}^{r}_{(\lambda,\mu,\alpha)}=
{}^{i}{\hat{O}}^{r_{1}r_{2}}_{\lambda, \mu}
\otimes{\hat{\sigma}}^{r_{3}}_{\alpha}.$ 
and calculate nonzero matrix element
\begin{equation}
\label{eq: R3.11}
<\lambda,\mu\mid {}^{i}{\hat{\gamma}}^{r}_{(\tau,\nu,\alpha)}\mid \tau,\nu>=
{}^{*}\delta_{\lambda\tau}{}^{*}\delta_{\mu,\nu}
{}^{i}e^{r}_{(\tau,\nu,\alpha)},
\end{equation}
where ${}^{i}e^{r}_{(\lambda,\mu,\alpha)}=
{}^{i}O^{r}_{\lambda,\mu} \otimes\sigma_{\alpha}$.
The set of operators $\left\{ 
{}^{i}{\hat{\gamma}}^{r} \right\}$
is the basis for all operator vectors 
$\hat{\Phi}(\zeta)={}^{i}{\hat{\gamma}}^{r}\,
{}^{i}\Phi_{r}(\zeta)$ of tangent section of 
principle bundle with the base of operator multimanifold
$\hat{G}_{N}=\left( \S_{i}^{N}\oplus {}^{*}\hat{\RR_{i}}^{4} \right)
\otimes\hat{R}^{3}$.
Here $^{*}\hat{\RR_{i}}^{4}$ is the $2\times2$ dimensional linear 
pseudo operator space, with the set of the linear unit operator 
pseudo vectors eq.(4.1.3) as the basis of tangent vector section,
and $\hat{R}^{3}$ is the three dimensional real linear operator space with
the basis consisted of the ordinary unit operator vectors  
$\{ {\hat{\sigma}}^{r}_{\alpha}\}$. 
The 
$\hat{G}_{N}$ is decomposed as follows:
\begin{equation}
\label{eq: R3.15}
\hat{G}_{N}=\hG_{\eta}\oplus\hG_{u_{1}}
\oplus \cdots \oplus\hG_{u_{N}},
\end{equation}
where $\hG_{u_{i}}$ is the six dimensional operator
manifold of the species $(i)$ with the basis \\
$\left\{ {}^{i}{\hgam_{u}}^{r}_{(\lambda\alpha)} =
{}^{i}{\hat{\O1_{u}}}_{\lambda}^{r}\otimes {\hat{\sigma}}^{r}_{\alpha}
\right\}$.
The expansions of operator vectors and 
covectors  are written 
$\hps_{\eta}={\hgam_{\eta}}^{r}
{\ps1_{\eta}}_{r}, \quad
\hps_{u}= {}^{i}{\hgam_{u}}^{r}\,
{}^{i}{\ps1_{u}}_{r}, 
\quad
\bar{\hps_{\eta}}=
{\hgam_{\eta}}_{r}
{\ps1_{\eta}}^{r},\quad
\bar{\hps_{u}}=
{}^{i}{\hgam_{u}}_{r}\,
{}^{i}{\ps1_{u}}^{r},$
where the components ${\ps1_{\eta}}_{r}(\eta)$ and
${}^{i}{\ps1_{u}}_{r}(u)$ are respectively the 
link functions of $\eta$-type and ${}^{i}u$-type structures.

\subsection{Field Aspect}
\label{Field}
The quantum field and differential geometric aspects of  $\hat{G}_{N}$
may be discussed on the analogy
of $\hat{G}_{N=1}$. Here we turn to some points of the field aspect. 
We consider the special system of 
regular structures, which is made of fundamental structure of $\eta$-type and 
infinite number of ${}^{i}u$-type ordinary structures of different species 
$(i=1,\ldots, N)$. 
To become stable the  primordial structures in this system 
establish a stable linkage
\begin{equation}
\label{eq: R4.1}
p^{2}=p^{2}_{\eta}- \S^{N}_{i=1}p^{2}_{u_{i}}=0.
\end{equation}
The free field defined on
multimanifold $G_{N}=\G1_{\eta}\oplus
\G1_{u_{1}}\oplus \cdots\oplus\G1_{u_{N}}$ is written
$$
\Psi =\ps1_{\eta}(\eta)\ps1_{u}(u),\quad 
\ps1_{u}(u)=\ps1_{u_{1}}(u_{1})\cdots\ps1_{u_{N}}(u_{N}), 
$$
where $\ps1_{u_{i}}$ is  the bispinor
defined on the internal manifold $\G1_{u_{i}}$.
A Lagrangian of free field reads
\begin{equation}
\label{eq: RC.22.1}
\widetilde{L}_{0}(D)=
\FFr{i}{2} \{ \bar{\Psi}_{e}(\zeta)
{}^{i}\gamma^{(\lambda,\mu,\alpha)}
{\pr_{i}}_{(\lambda,\mu,\alpha)}\Psi_{e}(\zeta)-
{\pr_{i}}_{(\lambda,\mu,\alpha)}\bar{\Psi}_{e}(\zeta)
{}^{i}\gamma^{(\lambda,\mu,\alpha)}
\Psi_{e}(\zeta) \}.
\end{equation}
We adopt the following conventions:
\begin{equation}
\label{eq: RC.22.2}
\begin{array}{l}
\Psi_{e}(\zeta)=e\otimes\Psi(\zeta)=
\left( \matrix{
1 &1\cr
1 &1\cr
} 
\right)  
\otimes\Psi(\zeta), 
\quad
\bar{\Psi}_{e}(\zeta)=e\otimes \bar{\Psi}(\zeta),\quad
\bar{\Psi}(\zeta)=\Psi^{+}(\zeta)\gamma^{0}, \\ 
{}^{i}\gamma^{(\lambda,\mu,\alpha)}=
{}^{i}{\widetilde{O}}^{\lambda,\mu}
\otimes{\widetilde{\sigma}}^{\alpha}, \quad
{}^{i}{\widetilde{O}}^{\lambda,\mu}=\FFr{1}{\sqrt{2}}
\left( \nu_{i}\xi_{0}\otimes
{\widetilde{O}}^{\mu}+\varepsilon_{\lambda}\xi \otimes{}^{i}
{\widetilde{O}}^{\mu}\right),\\ 
\varepsilon_{\lambda}=\left\{ \matrix{
1 &\lambda=1\cr
-1 &\lambda=2\cr
}\right., \quad
<\nu_{i},\nu_{j}>=\delta_{ij},\quad
\left\{ {}^{i} {\widetilde{O}}^{\lambda},{}^{j}{\widetilde{O}}^{\mu}
\right\}=\delta_{ij}{}^{*}\delta^{\lambda\mu},\\ 
{\widetilde{O}}^{\mu}=\FFr{1}{\sqrt{2}}
\left( \xi_{0}+
\varepsilon_{\mu}\xi \right),\quad
{\widetilde{O}}^{\lambda}=
{}^{*}\delta^{\lambda\mu}{\widetilde{O}}_{\mu}=
{({\widetilde{O}}_{\lambda})}^{+},\quad 
{}^{i}{\widetilde{O}}^{\mu}=\FFr{1}{\sqrt{2}}
\left( \xi_{0i}+
\varepsilon_{\mu}\xi_{i} \right),
\\
{\pr_{i}}_{(\lambda,\mu,\alpha)}=\partial/\partial\,{}^{i}\zeta^
{(\lambda,\mu,\alpha)}, \quad 
\xi_{0}=\left( \matrix{
1 &0 \cr
0 &-1\cr
}\right)
\quad
\xi=\left( \matrix{
0 &1 \cr
-1 &0\cr
}\right),\\ 
{\xi_{0}}^{2}=-\xi^{2}=-{\xi_{0i}}^{2}=\xi^{2}_{i}=1,\quad
\{\xi_{0},\xi\}=\{\xi_{0},\xi_{0i}\}=\{\xi_{0},\xi_{i}\}=\\ 
=\{\xi,\xi_{0i}\}=\{\xi,\xi_{i}\}=\{\xi_{0i},\xi_{j}\}_{i\neq j}=
\{\xi_{0i},\xi_{0j}\}_{i\neq j}=\{\xi_{i},\xi_{j}\}_{i\neq j}=0.
\end{array}
\end{equation}
Field equations are written
\begin{equation}
\label{eq: RC.22.3}
\begin{array}{l}
(\hp1_{\eta}-m)\ps1_{\eta}(\eta)=0,\quad       
\bar{\ps1_{\eta}}(\eta)(\hp1_{\eta}-m)=0,\\ 
(\hp1_{u}-m)\ps1_{u}(u)=0, \quad \bar{\ps1_{u}}(u)(\hp1_{u}-m)=0,
\end{array}
\end{equation}
where
\begin{equation}
\label{eq: RC.22.4}
\begin{array}{ll}
\hp1_{\eta}=i\hpr_{\eta}, \quad \hp1_{u}=i\hpr_{u},\quad 
\hpr_{u}={}^{i}\gamma^{(\lambda\alpha)}{\pr_{u_{i}}}_{(\lambda\alpha)},\quad
{\pr_{\eta}}_{(\lambda\alpha)}=\partial/\partial\eta^{(\lambda\alpha)},
\quad{\pr_{u_{i}}}_{(\lambda\alpha)}=\partial /
\partial u^{(\lambda\alpha)}_{i},\\ 
{}^{i}{\gam_{\eta}}^{(\lambda\alpha)}=
{}^{i}{\widetilde{\O1_{\eta}}}^{\lambda}\otimes
{\widetilde{\sigma}}^{\alpha}=
\nu_{i}\xi_{0}\otimes\gamma^{(\lambda\alpha)}=
\nu_{i}\xi_{0}\otimes{\widetilde{O}}^{\lambda}\otimes
{\widetilde{\sigma}}^{\alpha},\\ 
{}^{i}{\gam_{u}}^{(\lambda\alpha)}=
{}^{i}{\widetilde{\O1_{u}}}^{\lambda}\otimes
{\widetilde{\sigma}}^{\alpha}=
\xi\otimes{}^{i}\gamma^{(\lambda\alpha)}=
\xi\otimes{}^{i}{\widetilde{O}}^{\lambda}\otimes
{\widetilde{\sigma}}^{\alpha},\\ 
\left(\gamma^{(\lambda\alpha)}\right)^{+}=
{}^{*}\delta^{\lambda\tau}\delta^{\alpha\beta}\gamma^{(\tau\beta)}=
\gamma_{(\lambda\alpha)},\quad 
\left({}^{i}{\gam_{u}}^{(\lambda\alpha)}\right)^{+}=
-{}^{i}{\gam_{u}}_{(\lambda\alpha)}.
\end{array}
\end{equation}
The state of free ordinary structure of ${}^{i}u$-type with the given values 
of link momentum $\p1_{u_{i}}$ and spin projection $s_{i}$ is described 
by means of plane wave.
It is also necessary to consider the solution of negative
frequencies with the normalized bispinor amplitude.

\subsection {Realization of Multimanifold $G_{N}$}
\label {manif}
We consider a special stable system eq.(4.2.1).
In analogy with subsec.2.1  
we make use of localized wave packets by
means of superposition of plane wave solutions furnished 
by creation and annihilation operators in agreement with Pauli's 
principle.
Straightforward calculation gives the relation
\begin{equation}
\label{eq: R4.7}
\begin{array}{l}
\S_{\lambda=\pm}<\chi_{\lambda}\mid\hat{\Phi}(\zeta)
\bar{\hat{\Phi}}(\zeta)\mid
\chi_{\lambda}>= 
\S_{\lambda=\pm}<\chi_{\lambda}\mid
\bar{\hat{\Phi}}(\zeta)\hat{\Phi}(\zeta)\mid\chi_{\lambda}>= \\ 
=-i\Lm_{\zeta\rightarrow\zeta'}(\zeta\zeta')\G1_{\zeta}(\zeta-\zeta')=
-i\Lm_{\eta\rightarrow\eta'}(\eta\eta')\G1_{\eta}(\eta-\eta')-
i\Lm_{u_{i}\rightarrow u'_{i}}\S_{i=1}^{N}
(u_{i}u'_{i})\G1_{u_{i}}(u_{i}-u'_{i}),
\end{array}
\end{equation}
provided by the Green's function 
$
\G1_{u_{i}}(u_{i}-u'_{i})=-(i\hpr_{u_{i}}+m)\Dlt_{u_{i}}(u_{i}-u'_{i}),
$
where the  
$\Dlt_{u_{i}}(u_{i}-u'_{i})$
is the invariant singular function.
Thus
\begin{equation}
\label{eq: R4.9}
\zeta^{2}{\G1_{\zeta}}_{F}(0) = \eta^{2}{\G1_{\eta}}_{F}(0)-
\S^{N}_{i=1}{u_{i}}^{2}{\G1_{u_{i}}}_{F}(0),
\end{equation}
where ${\G1_{\eta}}_{F},
{\G1_{u}}_{F}$ and ${\G1_{\zeta}}_{F}$ are causal Green's functions
of the $\eta-,u-$ and $\zeta$-type structures.
The realization of the multimanifold stems from the condition
imposed  upon the matrix element eq.(4.3.1),
that as the bilinear form on operator vectors it is 
required to be finite
\begin{equation}
\label{eq: R4.10}
\S_{\lambda=\pm}<\chi_{\lambda}\mid\hat{\Phi}(\zeta)
\bar{\hat{\Phi}}(\zeta)\mid
\chi_{\lambda}>=
\zeta^{2}{\G1_{\zeta}}_{F}(0) < \infty.
\end{equation}
Let denote
$
u^{2}\G1_{u}(0)\equiv
\Lm_{u_{i}\rightarrow u'_{i}}\S_{i=1}^{N}(u_{i}u'_{i})
\G1_{u_{i}}(u_{i}-u'_{i})
$
and consider a stable system eq.(4.2.1). Hence
\begin{equation}
\label{eq: R4.12}
{\G1_{u}}_{F}(0)=
{\G1_{\eta}}_{F}(0)=
{\G1_{\zeta}}_{F}(0),
\end{equation}
provided
$m\equiv \left| p_{u}\right|=\left( \S_{i=1}^{N}
{p_{u_{i}}}^{2} \right)^{1/2}=\left| p_{\eta}\right|.$
According to eq.(4.3.5) and eq.(4.3.4),
the length of each vector
${\bf \zeta}={}^{i}e\,
{}^{i}\zeta\in G_{N}$
should be equaled zero
$\zeta^{2}=\eta^{2}-u^{2}=\eta^{2}-\S_{i=1}^{N}({u^{G}_{i}})^{2}=0,$
where use is made of
$$
u^{G}_{i}\equiv u_{i}
\left[ 
\Lm_{u_{i}\rightarrow u'_{i}}{\G1_{u_{i}}}_{F}(u_{i}-u'_{i})
\left. \right/
\Lm_{\eta\rightarrow\eta'}{\G1_{\eta}}_{F}(\eta-\eta')
\right]^{1/2},
$$
and 
$u^{G}_{i}={}^{i}{\he_{u}}_{(\lambda,\alpha)}
u_{i}^{G(\lambda,\alpha)}$.
Thus, the multimanifold $G_{N}$ comes into being, which is
decomposed as follows:
\begin{equation}
\label{eq: R4.18}
G_{N}=\G1_{\eta}\oplus\G1_{u^{G}_{1}}\oplus\cdots
\oplus \G1_{u^{G}_{N}}.
\end{equation}
It brings us to the conclusion: the major requirement eq.(4.3.3)
provided by stability condition eq.(4.3.4) or eq.(4.2.1) yields 
the flat multimanifold $G_{N}$.
Meanwhile the  Minkowski flat space $M^{4}$ stems from the
flat submanifold $\G1_{\eta}$ (subsec. 2.1), in which the line 
element turned out to be invariant. That is,
the principle of Relativity comes into being with $M^{4}$ ensued from
the multiworld geometry $G_{N}$.
In the subsequent paper (Part II) we shall use a notion of $i$-th 
internal world for the submanifold $\G1_{u_{i}}$.

\section{Concluding Remarks}
\label{Conc}

Our aim is to develop the operator manifold  formalism, which is
the mathematical basis for the presented approach to describe the
microscopic structures of particles (Part II).
It is a generalization of secondary quantization method with 
appropriate expansion over the geometric objects leading to the 
quantization of geometry, different from all existing
schemes. 
Based on configuration space mechanics with antisymmetric state functions,
we discuss in detail the quantum field and differential geometric aspects 
of the method of operator manifold. 
We develop the formalism of operator multimanifold yielding the
multiworld geometry. The value of the present version of hypothesis of 
existence of multiworld structures resides in solving some key problems 
of particle physics (Part II).

\centerline {\bf\large Acknowledgements}
\vskip 0.1\baselineskip
\noindent
I am pleased to mention the most valuable discussions with S.P.Novikov and
the late V.Ambartsumian on the various issues treated in this paper.
I express my gratitude to G.Jona-Lasinio for fruitful comments and suggestions.
I'm indebted to V.Gurzadyan, A.M.Vardanian and K.L.Yerknapetian
for support.

\section * {Appendix}
\subsection * {The Solution of Wave Equation of Distorted Structure}
\label {app}
\renewcommand {\theequation}{A.\arabic {equation}}
To solve the equation (3.3.13)
\begin{equation}
\label {eq: RB.18}
\left[ i g ({\pr_{u}}-\Gamma )- m \right]\Psi(u) =0,
\end{equation}
we transform it into
\begin{equation}
\label{R144}
\{-\partial^{2}-m^{2}-{(g\Gamma)}^{2}+2(\Gamma g)+
(g \partial)(g\Gamma) \}\Psi=0,
\end{equation}
where we abbreviate the
indices $(\lambda\alpha)$ by the single symbol $\mu$, and Latin indices
$(im)\quad (i=\pm, m=1,2,3)$ by $i$, also denote 
${\hat{\p1}_{u}}\equiv 
\hat{p}$ and
\begin{equation}
\label{R145}
\begin{array}{l}
\FFr{1}{2}\sigma^{\mu\nu}F_{\mu\nu}=(g\partial)(g\Gamma)-
(\partial\Gamma),\quad (g\partial)(g\Gamma)=
g^{\mu}g^{\nu}\partial_{\mu}\Gamma_{\nu}, \\
\FFr{1}{2}\sigma^{\mu\nu}[\Gamma_{\mu},\Gamma_{\nu} ]=
{(g\Gamma)}^{2}- \Gamma^{2}, \quad
\partial^{2}=\partial^{\mu}\partial_{\mu}, \quad
\Gamma^{2}=\Gamma^{\mu}\Gamma_{\mu}, \\
2g^{\mu\nu}=\{ g^{\mu},g^{\nu} \},\quad
2\sigma^{\mu\nu}= [g^{\mu},g^{\nu} ],\quad
\F_{\mu\nu}=\partial_{\mu}\Gamma_{\nu}-
\partial_{\nu}\Gamma_{\mu}.
\end{array}
\end{equation}
We are looking for a solution given in the form
$\Psi=e^{-ipu}F(\varphi)$,
where $p_{\mu}$ is a constant sixvector $pu=p_{\mu}u_{\mu}$,
and admit that the field of distortion is switched on
at $u_{0}=-\infty $ smoothly. Then the function $\Psi$ must match
onto the wave function of ordinary regular structure. 
Smoothness requires that the numbers $p_{\mu}$ become the components of 
link momentum of regular structure and satisfy the boundary condition
$p_{\mu}p_{\mu}=m^{2}=p^{2}_{\eta}$ eq.(3.1.3).
Due to it we cancel unwanted solutions
and clear up the normalization of wave functions 
\begin{equation}
\label{R146}
\int\Psi^{*}_{p'}\Psi_{p}d^{3}u=\int{\bar{\Psi}}_{p'}\gamma^{0}\Psi_{p}
d^{3}u={(2\pi)}^{3}\delta(\vec{p'}-\vec{p}).
\end{equation}
We suppose that at $\sqrt{-g}\neq1 $ the gradient of the function 
$\varphi $ reads
$$\partial_{\mu}\varphi=V^{i}_{\mu}k_{i}, \quad
\partial^{\mu}\varphi=V_{i}^{\mu}k^{i},$$
where $k_{i}$ are arbitrary constant numbers satisfying the condition
$k_{i}k_{i}=0$. Thus
$\partial^{\mu}\varphi\partial_{\mu}\varphi=
(V_{i}^{\mu}V^{j}_{\mu})k^{i}k_{j}=0$.
Then the eq.(14.4) gives rise to $F'=A(\theta)F$,
where $(\cdots)'$ stands for the derivative with respect to $\varphi$, and
\begin{equation}
\label{R147}
A(\theta)=\FFr{2i(p\Gamma) + m^{2}-p^{2}+{(g\Gamma)}^{2}-
(g\partial)(g\Gamma)}{2i(kVp)-(kDV)};
\end{equation}
where
\begin{equation}
\label{R148}
\begin{array}{ll}
(kVp)=k^{i}V_{i}^{\mu}p_{\mu},\quad
(kDV)=k^{i}D_{\mu}V_{i}^{\mu}, 
\quad
D_{\mu}=\partial_{\mu}-2\Gamma_{\mu}, \\
p^{2}=p^{\mu}p_{\mu}=g^{\mu\nu}(\theta)p_{\mu}p_{\nu}\quad
(kVdu)=k_{i}V^{i}_{\mu}du^{\mu}.
\end{array}
\end{equation}
We are interested in the right-handed eigenvectors $F_{r}\quad
(r=1,2,3,4)$ corresponding to eigenvalues $\mu_{r}$ of matrix $A:
AF_{r}=\mu_{r}F_{r}$, which are the roots of polynomial characteristic
equation 
$$
c(\mu)=\left\|(\mu I-A)\right\|=0.
$$
Thus, one gets $F'_{r}=\mu_{r}F_{r}$ and
$F=\displaystyle \prod^{4}_{r=1}F_{r}$.
Hence
$(\ln F)'=\S^{4}_{r=1}\mu_{r}=trA$
and $(\ln F)'=X_{R}(\theta)-iX_{J}(\theta)$,
provided
\begin{equation}
\label{R149}
\begin{array}{l}
X_{R}(\theta)=
trA_{R}(\theta)=tr \left\{ \FFr{-(kDV)\left[ m^{2}-p^{2}+
{(g\Gamma)}^{2}-(g\partial)(g\Gamma)\right]+4(kVp)
(p\Gamma)}{(kDV)^{2}+4(kVp)^{2}} \right\},\\
X_{J}(\theta)=
trA_{J}(\theta)=2tr \left\{ \FFr{(kVp)\left[ m^{2}-p^{2}+
{(g\Gamma)}^{2}-
(g\partial)(g\Gamma)\right]+(kDV)(p\Gamma)}{(kDV)^{2}+
4(kVp)^{2}} \right\}.
\end{array}
\end{equation}
The solution of eq.(A.1) reads
\begin{equation}
\label{R1410}
F(\theta)=C{\left( \frac{m}{E_{u}} \right)}^{1/2} U\exp
\{\chi_{R}(\theta)-i\chi_{J}(\theta)\}, 
\end{equation}
where $C=1$ is the normalization constant, $U$ is the constant bispinor, and
\begin{equation}
\label{R1410}
\chi_{R}(\theta)=\int_{0}^{u^{\mu}}(kVdu)X_{R}(\theta),\quad
\chi_{J}(\theta)=\int_{0}^{u^{\mu}}(kVdu)X_{J}(\theta).
\end{equation}

\end{document}